\documentclass[twocolumn, letterpaper, aps, prapplied, floatfix, superscriptaddress, 10pt]{revtex4-2}

\pdfoutput=1

\usepackage{amsmath}
\usepackage{amsthm}
\usepackage{amssymb}
\usepackage{booktabs}
\usepackage{graphicx}
\usepackage[abbreviations]{foreign}
\usepackage[colorlinks=true, linkcolor=blue, citecolor=blue, urlcolor=blue, pdfauthor={Kieran Dalton}, pdftitle={Resource-Efficient Cross-Platform Verification with Modular Superconducting Devices}]{hyperref}

\usepackage{mathtools}
\usepackage{mhchem}
\usepackage{orcidlink}
\usepackage{physics2}
\usepackage{siunitx}
\usepackage{soul} 
\usepackage{xcolor}

\DeclareMathOperator{\tr}{tr}

\newcommand{\BBM}{Bell-basis measurements}
\newcommand{\BBMprotocol}{Bell-basis measurement protocol}

\newcommand{\QST}{quantum state tomography}

\newcommand{\RM}{randomized measurements}

\DeclareSIUnit\mK{mK}
\DeclareSIUnit\sccm{SCCM}
\DeclareSIUnit\fluxquantum{\ensuremath{\Phi_0}}

\definecolor{todocolor}{HTML}{8c000f}

\usephysicsmodule{braket}

\newcommand{\ghzn}{\ket{\mathrm{GHZ}}_n}

\newcommand{\affiliationethzphys}{\affiliation{Department of Physics, ETH Zürich, CH-8093 Zurich, Switzerland}}
\newcommand{\affiliationethzqc}{\affiliation{Quantum Center, ETH Zürich, CH-8093 Zurich, Switzerland}}
\newcommand{\affiliationpsiqchub}{\affiliation{ETH Zürich---PSI Quantum Computing Hub, Paul Scherrer Institut, CH-5232 Villigen, Switzerland}}
 
\begin{document}
\title{Resource-Efficient Cross-Platform Verification\\ with Modular Superconducting Devices}

\author{Kieran Dalton\,\orcidlink{0009-0006-2416-356X}}
\email{kieran.dalton@phys.ethz.ch}
\affiliationethzphys{}
\affiliationethzqc{}
\affiliationpsiqchub{}

\author{Johannes Kn{\"o}rzer\,\orcidlink{0000-0002-7318-3018}}
\affiliationethzphys{}
\affiliationethzqc{}

\author{Finn Hoehne}
\affiliationethzphys{}

\author{Yongxin Song}
\affiliationethzphys{}
\affiliationethzqc{}
\affiliationpsiqchub{}

\author{Alexander Flasby}
\affiliationethzphys{}
\affiliationpsiqchub{}

\author{Dante \surname{Colao~Zanuz}}
\affiliationethzphys{}
\affiliationethzqc{}

\author{Mohsen Bahrami Panah}
\affiliationethzphys{}
\affiliationpsiqchub{}

\author{Ilya Besedin}
\affiliationethzphys{}
\affiliationethzqc{}
\affiliationpsiqchub{}

\author{Jean-Claude Besse\,\orcidlink{0000-0002-1490-0072}}
\affiliationethzphys{}
\affiliationethzqc{}
\affiliationpsiqchub{}

\author{Andreas Wallraff\,\orcidlink{0000-0002-3476-4485}}
\email{andreas.wallraff@phys.ethz.ch}
\affiliationethzphys{}
\affiliationethzqc{}
\affiliationpsiqchub{}

\date{\today}

\begin{abstract}
Large-scale quantum computers are expected to benefit from modular architectures.
Validating the capabilities of modular devices requires benchmarking strategies that assess performance within and between modules. 
In this work, we evaluate cross-platform verification protocols, which are critical for quantifying how accurately different modules prepare the same quantum state --- a key requirement for modular scalability and system-wide consistency. 
We demonstrate these algorithms using a six-qubit flip-chip superconducting quantum device consisting of two three-qubit modules on a single carrier chip, with connectivity for intra- and inter-module entanglement. 
We examine how the resource requirements of protocols relying solely on classical communication between modules scale exponentially with qubit number, and demonstrate that introducing an inter-module two-qubit gate enables sub-exponential scaling in cross-platform verification.
This approach reduces the number of repetitions required by a factor of four for three-qubit states, with greater reductions projected for larger and higher-fidelity devices. 
\end{abstract}

\maketitle

\section{Introduction}

The advent of quantum computation promises to accelerate the simulation of complex quantum systems and is expected to access a strictly larger complexity class of problems than those efficiently addressable with conventional computation \cite{Watrous2008, Aaronson2013b, Nielsen2010}. 
To achieve this goal, we require devices containing thousands of qubits capable of applying millions of gates without errors, with additional qubit number overhead from quantum error correction \cite{Dalton2024, Gidney2025b, Dalzell2025, Preskill2025}.
To reach this scale within practical constraints, it may be necessary to use modular devices consisting of multiple independent quantum processors linked with quantum channels \cite{Awschalom2021}. 
In the case of superconducting qubits, this modularity may be achieved between different modules in the same package \cite{Gold2021, Field2024, Wu2024e}, different packages in the same cryostat \cite{Burkhart2021, Niu2023}, and even different cryostats, either with native microwave interconnects \cite{Magnard2020, Storz2023} or transduction and optical interconnects \cite{Andrews2014, Mirhosseini2020, Han2021f, Sahu2022, Ang2024}.

As modular devices scale to larger qubit numbers, benchmarking techniques must also scale efficiently to remain practical. 
Existing strategies typically fall into two categories: component-level benchmarking, which assesses the performance of individual operations such as single- and two-qubit gates \cite{Magesan2012, Magesan2012a, Heya2025, Hashim2024a}, and holistic benchmarking, which evaluates the behavior of an entire device using metrics like quantum volume \cite{Cross2018} or mirror-circuit fidelity \cite{Proctor2022}.
While component-level approaches require extensive testing across all subsystems, holistic benchmarks often depend on classical simulation for result verification --- a step that becomes computationally prohibitive as system size increases \cite{Hashim2024a, Proctor2025}. 
This scalability constraint highlights the need for verification protocols that are both experimentally feasible and capable of capturing full-system performance in large, modular architectures. 

A promising approach to scalable, end-to-end benchmarking is cross-platform verification via distributed inner product estimation \cite{Elben2019, Anshu2022, Knoerzer2023, Denzler2025}. 
Here, one prepares a quantum state on one subsystem, that may be the result of a computation, and then estimates the inner product between this state and a nominally identical state prepared on another subsystem. 
For states with low magic (\emph{i.e.}, flat Pauli distributions) and low entanglement, this estimation can be carried out efficiently using only local operations and classical communication (LOCC) between the modules \cite{Hinsche2024}. 
However, for arbitrary mixed states, the number of repetitions required under LOCC scales exponentially with system size \cite{Anshu2022}. 
This limitation is lifted when modules are connected through quantum links, as will be necessary for executing quantum algorithms distributed over different modules. The possibility of quantum communication enables sample-efficient protocols such as the SWAP test or Bell-basis measurements~\cite{Buhrman2001,Ekert2002,GarciaEscartin2013,Cincio2018}. 

Previous experimental work on cross-platform verification has focused exclusively on the LOCC variant, for both ion traps \cite{Elben2019} and between ion traps and superconducting circuits \cite{Zhu2022e, Zheng2024a}. 
While two-copy measurements have been used to extract information about entanglement in atomic systems~\cite{Daley2012,Islam2015,Bluvstein2022}, to date, no experiments have demonstrated distributed inner-product estimation using quantum communication between separate modules. 
In this work, we evaluate different cross-platform verification protocols using our six-qubit modular superconducting device, consisting of two modules each containing three qubits on the same carrier chip. 
We benchmark the performance of this device, and use it to compare the relative performance of protocols using LOCC and protocols which make use of the quantum links between the modules. We consider both the fidelities achieved and the resource scaling required for protocols verifying one-, two- and three-qubit states. 
        
\section{Device design and performance}
\label{sec:results:device}

As superconducting quantum devices increase in scale, they face limitations in routing, crosstalk \cite{Karamlou2024,Kosen2024} and fabrication yield \cite{Smith2022a}. 
To alleviate these limitations, the largest devices use 3D-integration technologies such as flip-chip bonding \cite{Kim2023b, Acharya2025, Gao2025}. 
Here, device components are distributed across the opposing faces of two vertically stacked chips. 
Modular devices promise to further improve device performance and fabrication yield using more than two chips \cite{Bravyi2022a}.

Recent realizations of modular devices have focused on implementing two-qubit gates between qubits on different modules \cite{Gold2021, Field2024, Wu2024e, Norris2025, Ihssen2025}. 
Building upon the inter-module two-qubit gate of Ref.~\cite{Norris2025}, we present a superconducting device consisting of six flux-tunable qubits divided between two three-qubit modules on a single carrier chip. 
The qubits are coupled (with fixed interaction strength) in a two-by-three ladder topology, with both inter-module and intra-module two-qubit gates implemented using nanosecond-timescale qubit frequency control with individual qubit flux lines \cite{Strauch2003,DiCarlo2009, Hellings2025}.
We present the measured device parameters in App.~\ref{sec:appendix:device}, and the measurement apparatus in App.~\ref{sec:appendix:measurementapparatus}.

\begin{figure}[t]
    \centering
    \includegraphics[width=70mm]{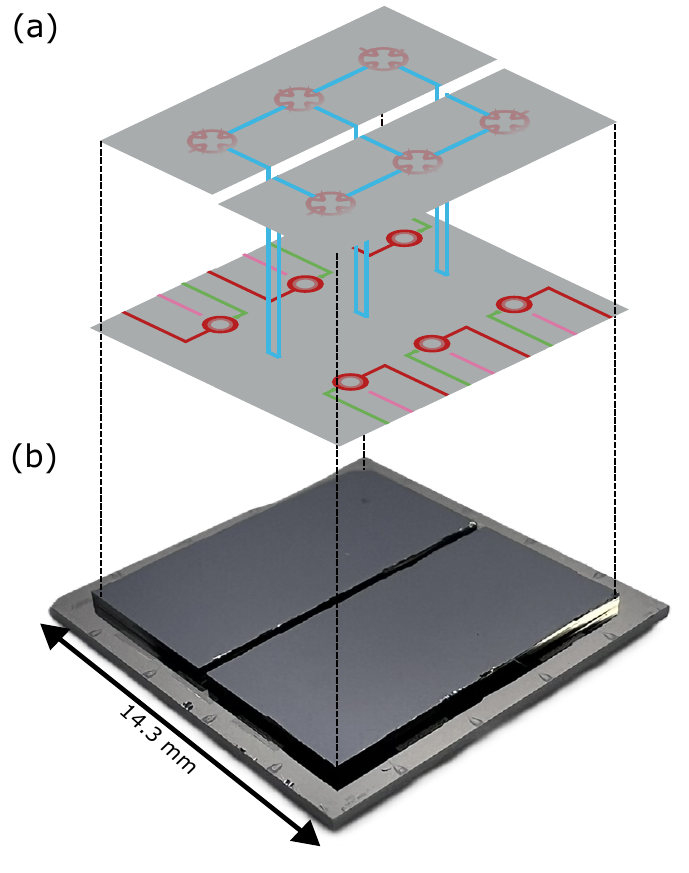}
    \caption{
        (a) Schematic of the device design showing the top modules and bottom carrier, with the qubit island (gaps where metal is removed are colored light red), qubit-qubit coupler (light blue), readout coupling pad (red), flux line (green), charge line (pink), and ground plane (gray). 
        The vertical separation between the chips is \qty{10}{\um}.
        (b) Photo of the assembled device. 
    }
    \label{fig:main:device}
\end{figure}

Each set of three qubits is read out using a frequency-multiplexed feedline coupled to individual Purcell filters attached to each qubit's readout resonator \cite{Walter2017, Heinsoo2018}. 
All readout circuitry and control lines are routed on the bottom carrier chip. 
When two lines must cross, one line is routed through indium bumps via the opposite chip. 
The two \qty{5.8}{\mm} by \qty{12}{\mm} modules are flip-chip bonded to the \qty{14.3}{\mm} by \qty{14.3}{\mm} carrier, see conceptual sketch in Fig.~\ref{fig:main:device}a and photograph of the assembled device in Fig.~\ref{fig:main:device}b. 
Further design considerations, and the device fabrication procedure, are described in Refs.~\cite{Norris2024, Norris2025}.

We characterize the device performance by benchmarking readout and single- and two-qubit gates.  
For qubit (qutrit) single-shot readout, we measure a median assignment error of \SI{1.01}{\percent} (\SI{2.21}{\percent}), with a range \SI{0.80}{\percent}\textendash\SI{1.43}{\percent} (\SI{1.77}{\percent}\textendash\SI{3.45}{\percent}).
For all measurements, we use single-shot readout both at the start and end of the circuit, with a pre-selection pulse to discard cases where residual excitation errors are detected. 
This rejects \SI{0.78}{\percent} (\SI{0.24}{\percent}\textendash\SI{2.76}{\percent}) of preparations per qubit. 
When preparing a three-qubit state, we obtain a total rejected fraction due to pre-selection of \SI{1.7}{\percent} (\SI{1.8}{\percent}) on the left (right) module, with differences from the individual qubit characterization due to fluctuating qubit relaxation times (see App.~\ref{sec:appendix:device}).
We benchmark single-qubit gates using randomized benchmarking \cite{Magesan2012a}, obtaining a median single-qubit error rate of \SI{0.11}{\percent} (\SI{0.06}{\percent}\textendash\SI{0.20}{\percent}).
Similarly, interleaved randomized benchmarking \cite{Magesan2012} provides a median two-qubit error rate of \SI{1.45}{\percent} (\SI{0.78}{\percent}\textendash\SI{1.56}{\percent}).
These are comparable to error rates achieved in similar planar \cite{Krinner2022, Besedin2025} and 3D-integrated \cite{Kosen2022,Field2024,Norris2025} devices. 
However, the two-qubit gate fidelities are lower than those in large 3D-integrated devices using tunable couplers \cite{Kim2023b, Acharya2025, Gao2025}. 
For all experiments, following data acquisition, we post-select to remove the repetitions where qubits are detected in the second excited state. 
We measure a median population leakage into the second excited state of \SI{0.21}{\percent} (\SI{0.10}{\percent}\textendash\SI{0.52}{\percent}) per two-qubit gate. 
This leakage is included in the aforementioned two-qubit error rates. 
When preparing a three-qubit state with two entangling gates, we obtain a total post-selected fraction due to leakage of \SI{0.49}{\percent} (\SI{0.70}{\percent}) on the left (right) module. 
We do not use constructive interference for leakage calibration \cite{Scarato2025} or leakage reduction pulses \cite{Miao2023a, Lacroix2025}.
We present the error rates for all qubits and two-qubit gates in a cumulative distribution plot, Fig.~\ref{fig:main:performance}a.

To carry out a holistic assessment of device performance, we prepare a maximally entangled three-qubit Greenberger-Horne-Zeilinger (GHZ) state on each three-qubit module. 
We use quantum state tomography (QST) \cite{Vogel1989} with maximum-likelihood estimation \cite{Neeley2010a} to find the most likely physical density matrix of the state. 
This involves measuring the state in an over-complete basis set of single-qubit Pauli bases (i.e. $X, Y$ and $Z$ bases) and their negatives $-X, -Y$ and $-Z$.
This allows us to correct for systematic errors in measurement axis rotations and detection bias \cite{Zhu2014a}. 
Additionally, we use calibration measurements of all qubits in their computational basis states for readout-error correction \cite{Steffen2006a}. 

The resulting three-qubit states have measured fidelities (according to Eq.~\ref{eq:def-fidelity}) of \qty{96.3}{\percent} and \qty{95.2}{\percent}. 
We simulate the state preparation circuit (the grey part of Fig.~\ref{fig:main:circuit}) using a realistic error model which takes into account the measured median error rates and relaxation times (see App.~\ref{sec:appendix:simulation}). The simulated fidelity is \qty{95.8}{\percent}, consistent with the measured fidelities. 
The infidelity is dominated by two-qubit gate errors. 
The Pauli basis decomposition of the resulting density matrix is presented in Fig.~\ref{fig:main:performance}b for the left module shown in Fig.~\ref{fig:main:device}(a,b). 

\section{Distributed Inner Product Estimation}
\label{sec:results:dipe}

\subsection{Quantum State Tomography}

Distributed inner product estimation is widely recognized as an important method for benchmarking near-term quantum computers~\cite{Elben2019,Anshu2022,Arunachalam2024,Zhu2022e, Hashim2024a, Proctor2025}. 
The task of distributed inner product estimation is to infer the inner product, $\tr(\rho_A \rho_B)$, between two spatially separated quantum states $\rho_A$ and $\rho_B$ with the same Hilbert space dimension, which are hosted in two different modules $A$ and $B$.
If the state purities are also known, knowledge of the inner product allows the state fidelity
\begin{equation}\label{eq:def-fidelity}
    \mathcal{F}(\rho_A, \rho_B) = \frac{\tr(\rho_A \rho_B)}{\max \{ \tr(\rho_A^2), \tr(\rho_B^2) \} }.
\end{equation}
\noindent to be calculated \cite{Jozsa1994}. If either state is pure, \emph{i.e.}, if $\tr(\rho_A^2) = 1$ or $\tr(\rho_B^2) = 1$, the inner product in the nominator coincides with the state fidelity.

\begin{figure*}[t]
    \centering
    \includegraphics[width=165mm]{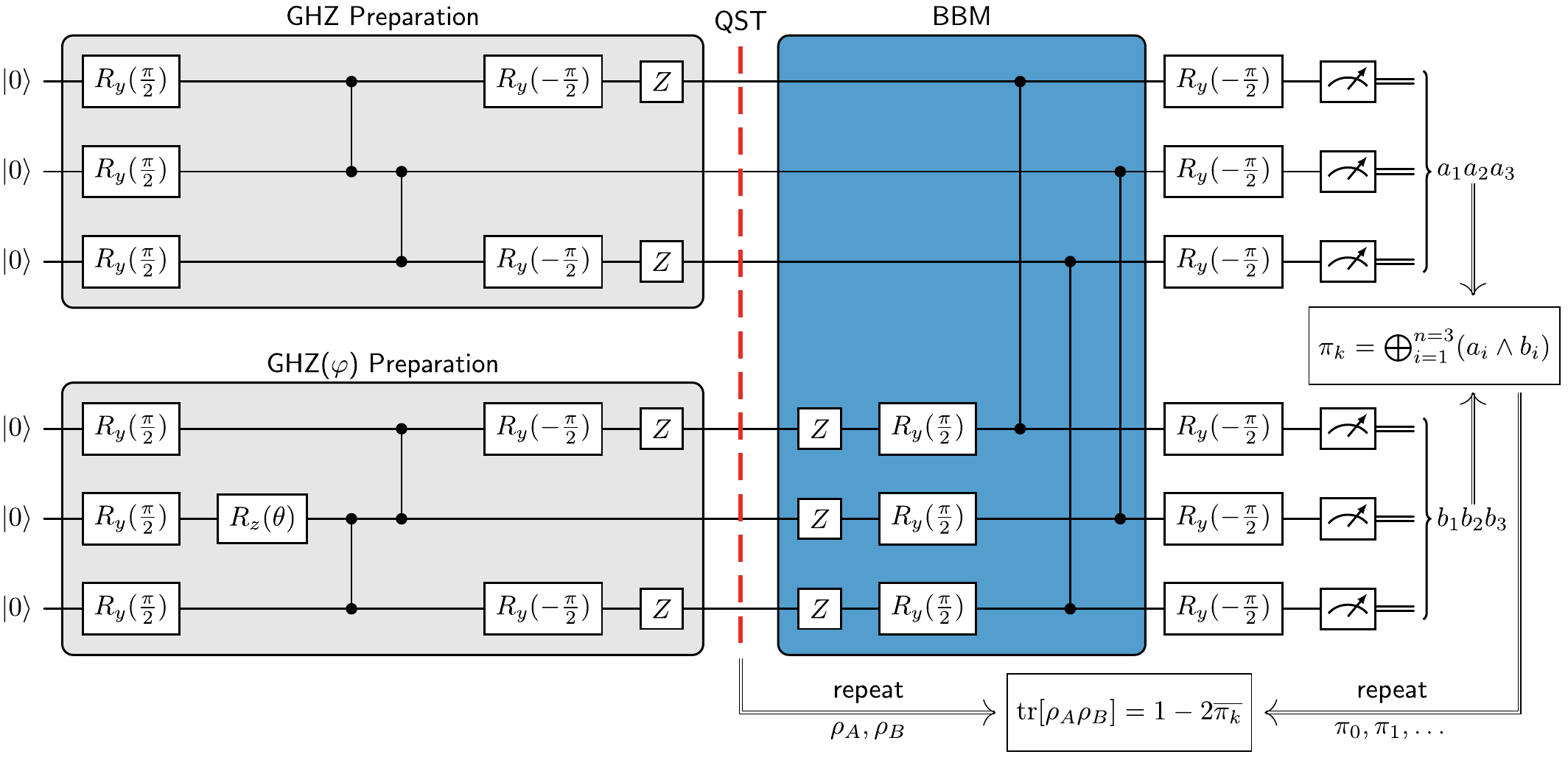}
    \caption{
        Circuit diagram for the GHZ state preparation and the Bell-basis measurement (BBM) protocol. 
        Measuring corresponding qubits on each module in the Bell basis results in two bitstrings $a_i$ and $b_i$. 
        Analyzing these as described in the main text results in an estimate of the inner product between the states on each module. 
        The red dotted line indicates where the quantum state tomography (QST) circuit would conclude, with measurements in a complete basis set.
    }
    \label{fig:main:circuit}
\end{figure*}

To benchmark the performance of both three-qubit modules, we initially prepare the phase-shifted $n$-qubit GHZ state on each module, defined as
\begin{equation}
\ghzn^{(\varphi)} = \frac{1}{\sqrt{2}} \left( \ket{0}^{\otimes n} + e^{i\varphi} \ket{1}^{\otimes n} \right)
\end{equation}
\noindent where $n \in \{1,2,3\}$.
On one of the modules, we prepare a GHZ state with $\varphi=0$.
On the other, we apply a variable phase $\varphi$ to the prepared state, as described with a conceptual circuit for $n=3$ in the gray region of Fig.~\ref{fig:main:circuit}. 
We vary this phase to test both nominally-identical and nominally-orthogonal states. 
In our implementation of this circuit, we take advantage of virtual-$Z$ gates \cite{McKay2017}. This results in the separable $2n$-qubit state $\ghzn^{(0)} \otimes \ghzn^{(\varphi)}$.

\begin{figure}[]
    \centering
    \includegraphics[width=75mm]{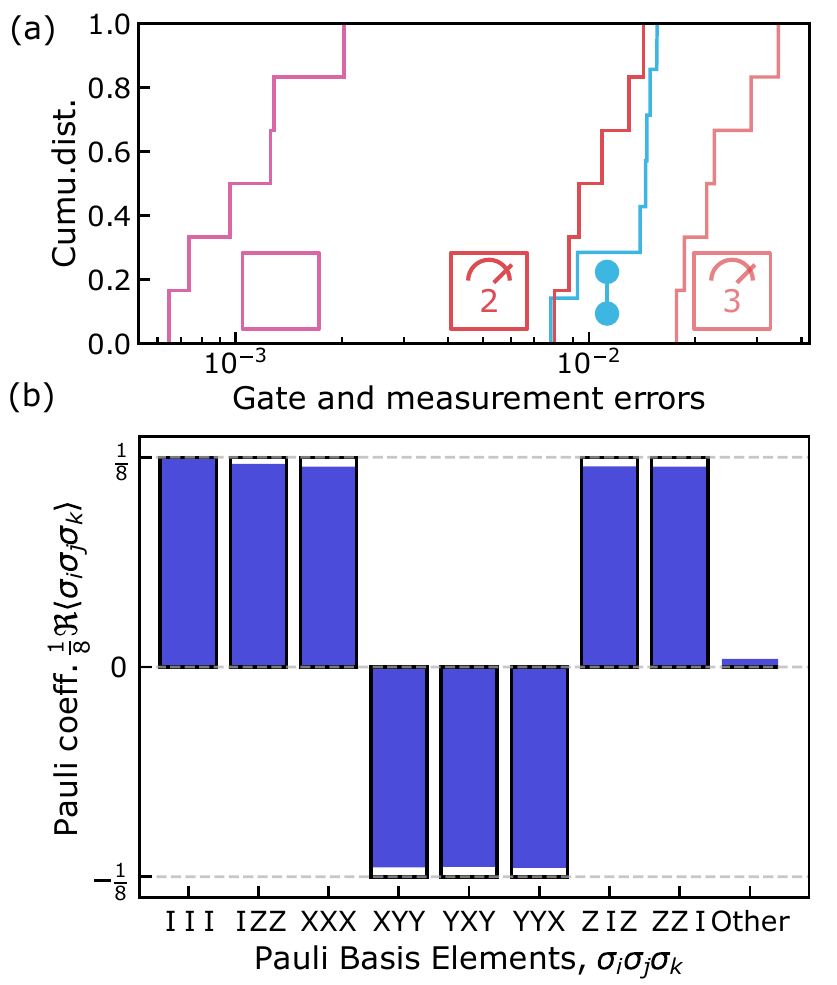}
    \caption{
        (a) Empirical cumulative distribution function of device errors, with the error rates of single-qubit gates (pink square), two-qubit gates (double cyan circles), and readout in the qubit and qutrit manifold (red measurement icons). 
        (b) State tomography result for a three-qubit GHZ state prepared on the three qubits of the left module. 
        We show Pauli decomposition coefficients which are expected to be non-zero. Ideal results in the absence of noise are represented by the black outlined bars.
        The remaining coefficient with the largest absolute value is presented as ``Other". 
    }
    \label{fig:main:performance}
\end{figure}

We carry out $n$-qubit quantum state tomography to estimate the density matrix for the single-module $n$-qubit subsystem $\ghzn^{(0)}$. 
We then repeat this on the other module to find the density matrix of the $n$-qubit subsystem $\ghzn^{(\varphi)}$, for 15 equally-spaced values of $\varphi \in \left[0, 2\pi\right]$. 
While only one GHZ state on one module is characterized, the full $2n$-qubit state is always prepared. 
This ensures that any inter-module crosstalk or spectator errors \cite{Krinner2020} due to the operations on one module affect the state characterized on the other module. 
For each \QST{} repetition, each set of bases and calibration points is measured $10^4$ times. 

We then calculate the inner product between the zero-phase state on the first module and each state of a given $\varphi$ prepared on the second module. 
For $n \in \{1,2,3\}$, the estimated inner product shows qualitative agreement with the calculated value of $\mathcal{F}(\ghzn^{(0)}, \ghzn^{(\varphi)}) = (1+\cos{\varphi})/2$, see Fig.~\ref{fig:main:cpv}(a,b), with near-unity inner product for identically-prepared states, and near-zero overlap for states with opposite phase. 
The peak measured inner products are lower for greater numbers of qubits, as one would expect from deeper state-preparation circuits. 

\subsection{Bell-Basis Measurements}
\label{sec:results:bbm}

If quantum information can be shared among the modules $A$ and $B$, it is possible to perform a SWAP-test~\cite{Buhrman2001} or Bell-basis measurements~\cite{GarciaEscartin2013,Cincio2018} between them to carry out distributed inner product estimation. \BBM{} consist of pair-wise two-qubit measurements in the Bell basis between corresponding qubits on each module. Although \BBM{} involve additional classical post-processing, they do not require an ancilla qubit, as the SWAP test does.

We now compare the inner products measured using quantum state tomography to those estimated with \BBM{}. 
In a separate measurement run, we prepare the same separable $2n$-qubit state, $\ghzn^{(0)} \otimes \ghzn^{(\varphi)}$, and then apply the \BBM{} circuit presented in the blue shaded region of Fig.~\ref{fig:main:circuit}.

Measuring all $2n$ qubits, one parameterized by $\varphi$ and one with $\varphi=0$, results in two bitstrings of length $n$: $a_i$ and $b_i$. 
The parity of the bitstring resulting from the bit-wise $AND$ operator ($\land$) between the two bitstrings is calculated, $\pi_k = \bigoplus_{i=1}^{n} \left( a_i \land b_i \right)$. 
The mean value of this parity, over many repetitions, determines the inner product, $\tr\left( \rho_A \rho_B \right) = 1 - 2 \overline{\pi_k}$. 
We repeat the algorithm for the same 15 values of $\varphi$, $10^4$ times each.

For \BBM{}, we again observe qualitative agreement with the calculation, with trace overlap less than \qty{0.1}{} for nominally-orthogonal states and greater than \qty{0.7}{} for nominally-identical states, qualitatively the same behavior as for \QST{}, see Fig.~\ref{fig:main:cpv}(c,d). 
The deviation from theory is greatest (above \qty{0.25}{} for $n=3$) for nominally-identical states with an expected trace overlap of one, and also high (above \qty{0.05}{} for $n=3$) for nominally-orthogonal states with zero expected trace overlap. 
This is because noise reduces the purity of the prepared GHZ states by degrading the coherent superpositions between $\ket{0}^{\otimes n}$ and $\ket{1}^{\otimes n}$.
By reducing state purity, the trace overlap $\tr (\rho_A \rho_B)$ for nominally-identical states is suppressed.
Similarly, for nominally-orthogonal states, noise enhances the trace overlap, as the different states tend towards the same maximally-mixed state (in the case of simple depolarizing noise). 

The inner product contrast is lower for the \BBM{} data than for the \QST{} data, mainly due to the additional two-qubit gates required by the \BBMprotocol{}. 
Additionally, the reliance of the \BBMprotocol{} on bitstring parity makes it more sensitive to readout misclassification errors. 
We expand on this further with an error-budget analysis in App.~\ref{sec:appendix:simulation}. 
Using prior measurements of a readout assignment matrix, the \BBM{} data can be readout-error corrected. 
This reduces the maximal deviation from theory to \qty{0.17}{} for $n=3$, closer to the \QST{} value of \qty{0.13}{}. 
Due to the additional measurements required, and device performance variations (see App.~\ref{sec:appendix:device}), we did not readout-error correct the \BBM{} data in Fig.~\ref{fig:main:cpv}(c,d).

\section{Sampling complexity analysis}
\label{sec:results:complexity}

\begin{figure*}[t!]
    \centering
    \includegraphics[width=165mm]{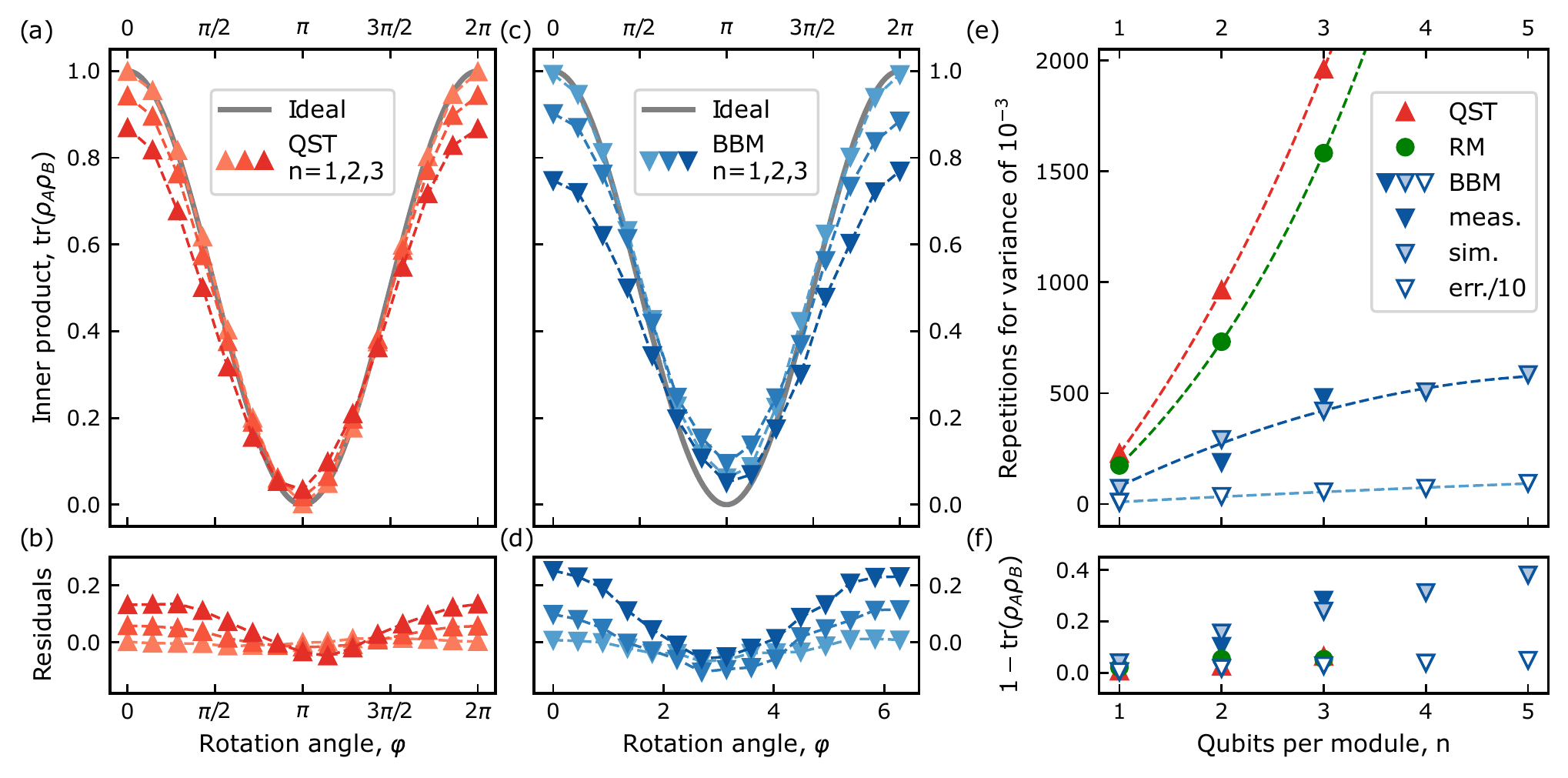}
    \caption{
        (a-d) Measured and calculated inner products for two $n$-qubit states, $\frac{1}{\sqrt{2}} \left( \ket{0}^{\otimes n} + \ket{1}^{\otimes n} \right)$ and $\frac{1}{\sqrt{2}} \left(\ket{0}^{\otimes n} + e^{i \varphi} \ket{1}^{\otimes n} \right)$, prepared on the left and right modules respectively. 
        The overlap is calculated using quantum state tomography (QST, red symbols) and Bell-basis measurements (BBM, blue symbols). 
        Statistical error bars are smaller than symbol sizes.
        (e) The number of measurements needed to estimate the inner product between two identically-prepared $\ghzn^{(0)}$ states to a statistical variance of $10^{-3}$. 
        The variance is calculated by taking 100 samples of varying size, with replacement, from a large dataset. 
        The relationship between variance and the sample size is then interpolated for the y-axis values. 
        For both plots, simulated data for the \BBMprotocol{}, using the median errors presented in Fig.~\ref{fig:main:performance}a, are presented as light blue downward triangles. White downward triangles represent simulations with all error rates divided by ten.
        Exponential (for \QST{} and \RM{}) and quadratic (for \BBM{}) fitted curves are provided as dashed lines as a guide to the eye. 
        Statistical error bars are smaller than symbol sizes.
        (f) The measured inner product for identically-prepared $\ghzn^{(0)}$ states using \BBM{}, \QST{}, and randomized measurements (RM).
    }
    \label{fig:main:cpv}
\end{figure*}

We now examine the relative scaling, with Hilbert space dimension, of the number of repetitions required for distributed inner product estimation, with and without quantum communication between the modules. 
We record a large dataset of $10^4$ repetitions (for each basis) for \QST{} and $1.5 \times 10^5$ repetitions for \BBM{}, and calculate the inner products of two identically-prepared $\ghzn^{(0)}$ states (for $n=1,2,3$) using both methods.

Quantum state tomography, particularly involving an overcomplete basis set, is inefficient for distributed inner product estimation and thus not scalable to large system sizes \cite{Anshu2022}. 
As a more efficient method, we use the same dataset to estimate the inner product using randomized measurements and classical correlations \cite{Elben2019,Elben2022}, selecting $3^n$ measurement bases using the greedy algorithm presented in App.~\ref{sec:appendix:rm}. 
We choose $3^n$ measurement settings as a balance between sample efficiency and the variance of the resulting inner product. 
Previous experimental demonstrations of cross-platform verification in ion-trap and superconducting devices used this randomized-measurement approach \cite{Elben2022, Zhu2022e}. 

We estimate the variance of the inner product for a given number of repetitions using bootstrapping.
Specifically, we draw $N_p$ repetitions from the dataset 100 times, sampling with replacement so that repetitions may appear multiple times.
For \QST{}, each selected repetition includes all measured bases.
We calculate the inner product for each of the 100 subsamples of size $N_{p}$, and determine the variance of these values.
For randomized measurements, we perform a new basis selection for each subsample.

We repeat this procedure for multiple subsample sizes $N_{p}$ and fit the resulting variances to a power law.
An example dataset is provided in App. Fig.~\ref{fig:appendix:sim_and_rm}b.
This fit allows us to determine the number of measurements required to reach a given target variance, which we choose to be $10^{-3}$.
We choose this value such that the statistical variations are lower than the systematic reductions of inner product due to noise, which are all greater than $10^{-2}$.
This method allows us to compare all three protocols on an even footing.

For LOCC-based methods (\QST{} and \RM{}), we see the number of necessary repetitions rise quickly with system size, reaching over \qty{1500}{} for $3$-qubit states.
The data agree well with an exponential fit to the measured \QST{} (red triangles) and \RM{} (green circles) data, see the dashed lines in Fig.~\ref{fig:main:cpv}e. 
This aligns with theory: estimating the inner product $\tr (\rho_A \rho_B)$ using LOCC requires an exponential number of measurements~\cite{Anshu2022}.
If one first tomographically reconstructs both $\rho_A$ and $\rho_B$ to compute $\tr (\rho_A \rho_B)$, it is evident that the exponential scaling of resources is inherited from \QST{} \cite{Haeffner2005}, which is currently practically limited to approximately $10$-qubit states~\cite{Song2017}. 
Since only three system sizes are measured, the empirical range over which this scaling can be observed is limited.

\BBM{}, on the other hand, require significantly fewer repetitions for all $n$. For $n=3$, \BBM{} require four times fewer repetitions than \QST{} for inner product estimation with a variance of $10^{-3}$, see the dark blue triangles in Fig.~\ref{fig:main:cpv}e. 
However, the measured scaling for \BBM{} is not constant with the number of qubits, as expected theoretically in the absence of noise~\cite{Knoerzer2023}.
This can be explained by considering the increase in error rates with system dimension. 
As the number of qubits is increased, the measured inner product decreases, and hence the probability of an even-parity bitstring in \BBM{} analysis also decreases (provided the inner product exceeds \qty{0.5}{}). 
Considering each experimental repetition as a Bernoulli trial with a success probability close to unity, one would expect a linear increase in variance for a fixed number of repetitions, with a quadratic contribution for trace overlaps further from unity. 

We emphasize that the improved scaling of \BBM{} must be considered alongside its additional hardware requirements, particularly when the two-qubit gates in the Bell-basis measurements cannot be executed in a single parallel operation. 
To assess these hardware requirements further, we simulate the \BBMprotocol{} using a symmetric depolarizing error channel with the median gate and readout error rates presented in Fig.~\ref{fig:main:performance}a. 
We provide further details of these simulations in App.~\ref{sec:appendix:simulation}.
Performing the same analysis as for the experimental data, the simulated data agree quantitatively with the measurements, indicating a faithful simulation. 
These simulated data are represented by the light blue downward triangles in Fig.~\ref{fig:main:cpv}(e,f). 
Reducing all error rates by a factor of ten, our simulated data approach the expected constant scaling without noise, with near-unity trace overlaps (see the white downward triangles in Fig.~\ref{fig:main:cpv}(e,f)). 
We provide a quadratic fit to the simulated \BBM{} data as a guide to the eye. This closely follows the data, and has the theoretically-expected negative coefficient for the quadratic term.

We emphasize that the expected exponential scaling complexity for LOCC methods applies to generic, arbitrary quantum states. 
For specific classes of states or with prior knowledge, they can perform significantly better. 
For instance, efficient tomographic methods are known if $\rho$ is fairly pure~\cite{Gross2010a} or well-approximated by a small matrix-product state~\cite{Cramer2010}. 
Efficiency can also improve by replacing standard Pauli measurements with alternative bases such as mutually unbiased bases~\cite{Acharya2025b} or threshold \QST{}~\cite{Binosi2024}. 
Inner product estimation similarly benefits from prior knowledge:
for many pure target states $\rho_A$ that are known classically, the overlap $\tr(\rho_A \rho_B)$ with a prepared state $\rho_B$ can be estimated with a constant number of measurements~\cite{Flammia2011}. 
Even when both states are unknown, efficient estimation is possible for states with low magic (\emph{i.e.}, flat Pauli distributions \cite{Bartlett2014}) and low entanglement~\cite{Hinsche2024}.

\section{Discussion and Outlook}

In this work, we demonstrated efficient cross-platform verification using three inter-module coupling buses between two three-qubit modules on a single carrier chip. 
We compared the performance of the sample-efficient Bell-basis measurement protocol \cite{GarciaEscartin2013,Cincio2018} to methods using only local operations and classical communication (LOCC) between the modules. 
We compared them both in terms of the estimated inner product between identically prepared states on each module and the resource scaling with states of different dimension. 

We demonstrated that, although the \BBMprotocol{} introduces inter-module two-qubit gates that modestly reduce the measured inner product compared to LOCC-based protocols, \BBM{} require substantially fewer repetitions to achieve a fixed statistical precision for states of increasing dimension, in contrast to the exponential resource scaling of trace-overlap estimation based on quantum state tomography~\cite{Cramer2010} and randomized measurements~\cite{Elben2022}. 
Despite the small sample size, our data agrees well with exponential scaling for LOCC-based protocols, and quadratic scaling for \BBM{}. 
Notably, to estimate the trace overlap between two nominally-identical $3$-qubit GHZ states to a given variance, \BBM{} require four times fewer repetitions than \QST{}. 
This highlights the utility of quantum communication in distributed quantum architectures. We expect that the improved scaling will become more relevant as modular quantum systems grow in size. 
Provided two-qubit gate and readout errors are reduced, this establishes \BBM{} between modules as a valuable tool for the scalable benchmarking of modular quantum devices, a critical requirement as quantum processors grow beyond monolithic architectures.

While we demonstrated cross-platform verification for three-qubit GHZ states, future studies could explore more general scenarios, such as verifying transformations or channels, which requires adapting \BBM{} to compare unitaries or process matrices rather than states~\cite{Knoerzer2023}.
This, however, comes at a cost. 
Standard approaches such as \BBM{} or SWAP tests for process fidelity estimation involve averaging over Haar-random input states, which is challenging to implement in practice.
In App.~\ref{sec:appendix:unitaries}, we present a simplified alternative based on sampling input states from the computational basis, which is straightforward to realize experimentally.
Although the resulting measure is not a true process fidelity, since it lacks basis independence, it can still serve as a useful performance metric of unitary processes.

Another interesting direction is the study of hybrid verification protocols that combine the strengths of both \BBM{} and LOCC techniques.
For example, local operations may serve to diagnose local errors or reduce state space before \BBM{} are applied~\cite{Arunachalam2024}.
As quantum systems continue to scale, and modular architectures become increasingly prominent, such capabilities will be essential for the reliable certification and operation of distributed quantum processors.

In addition to its utility for cross-platform verification, the ability to perform scalable Bell-basis measurements between separate modules may also prove valuable for Bell-sampling tasks, in which two identical copies of a state are prepared by a quantum circuit and measured in the Bell basis.
The measurement results yield information about the underlying quantum state, as they allow to estimate all $4^n$ Pauli observables of a $n$-qubit state highly efficiently~\cite{Huang2021f} and diagnose circuit errors~\cite{Montanaro2017, Hangleiter2024a}.

\section{Data availability}

The data produced in this work are available from the corresponding author upon reasonable request.

\section{Acknowledgments}

We thank Graham J. Norris and Christoph Hellings for helpful discussions regarding the experimental setup, and the technical staff at the Quantum Device Laboratory for their support. 
We thank Andreas Elben for providing feedback on the manuscript, and Johannes Fink for useful discussions regarding optical interconnects and transduction.

The authors acknowledge financial support by the Swiss State Secretariat for Education, Research and Innovation under contract number UeM019-11, by the Intelligence Advanced Research Projects Activity (IARPA) and the Army Research Office, under the Entangled Logical Qubits program and Cooperative Agreement Number W911NF-23-2-0212, and by ETH Zurich.
The views and conclusions contained in this document are those of the authors and should not be interpreted as representing the official policies, either expressed or implied, of IARPA, the Army Research Office, or the U.S.\ Government.
The U.S.\ Government is authorized to reproduce and distribute reprints for Government purposes notwithstanding any copyright notation herein.

\section{Contributions}

J.-C.B., J.K. and K.D.\ planned the experiments.
K.D.\ designed and K.D., D.C.Z., A.F., and M.B.P.\ fabricated the device.
K.D.\ and F.H.\ performed all measurements and analyzed all data.
Y.S.\ and I.B.\ assisted with the experimental setup and measurements.
K.D.\ and J.K.\ wrote the manuscript with input from all authors. \\
J.-C.B.\ and A.W.\ supervised the work.

\appendix

\renewcommand{\thefigure}{A\arabic{figure}}
\setcounter{figure}{0}
\renewcommand{\thetable}{A.\arabic{table}}
\setcounter{table}{0}

\section{Device parameters}
    \label{sec:appendix:device}
    
    Here we present the device parameters of our six-qubit modular device. 
    We measure the extremal dressed qubit $\ket{\mathrm{g}}$-$\ket{\mathrm{e}}$ and $\ket{\mathrm{e}}$-$\ket{\mathrm{f}}$ transition frequencies from Ramsey measurements as a function of the flux bias applied to the qubit, and use these to extract the qubit anharmonicity. 
    These values are presented in the upper half of Table~\ref{tab:appendix:hamiltonian-parameters:device-a}.

    With these parameters, we then extract the qubit charging energy and the qubit Josephson energy by numerically diagonalizing the coupled qubit-resonator-Purcell filter Hamiltonian, as described in Ref.~\cite{Norris2025}. 
    These parameters are given in the lower half of Table~\ref{tab:appendix:hamiltonian-parameters:device-a}.

    Qubit relaxation and dephasing times are known to vary over time \cite{Burnett2019, Lisenfeld2023}. 
    To characterize this, we perform 20 measurements of the qubit longitudinal relaxation times ($T_{1}$) and transverse relaxation times ($T_{2}^{*}$) over a period of 25 hours. 
    The coefficients of variation (standard deviation divided by mean) for the measured $T_{1}$ times are $\{ 0.20, 0.11, 0.29, 0.15, 0.13, 0.10 \}$ for the six qubits on the device.
    For $T_{2}^{*}$ times, we measure $\{ 0.42, 0.36, 0.35, 0.28, 0.21, 0.25 \}$. 
    This high degree of variation leads to fluctuations in gate and readout fidelities, and hence state preparation fidelities. 
    We do not characterize these variations in this work.
    
    \begin{table}[t]
        \centering
        \caption{
            Measured and extracted qubit parameters. 
            Measured parameters:
            Dressed qubit $\ket{\mathrm{g}}$-$\ket{\mathrm{e}}$ transition frequency  $\omega_\mathrm{q,ge}^\mathrm{d}(\phi)$ at reduced flux $\phi = \Phi / \Phi_0$ (with $\Phi$ the flux threading the SQUID loop and $\Phi_0$ the flux quantum); qubit anharmonicity  $\omega_\mathrm{q,ge}^\mathrm{d}(\phi)$.
            Extracted parameters: Josephson energy at $\phi = 0$, $E_\mathrm{J,max}$; charging energy $E_\mathrm{c}$.
        }
        \label{tab:appendix:hamiltonian-parameters:device-a}
        \begin{tabular}{l S S S S} \toprule
            && \multicolumn{3}{c}{{Left Module Qubits}} \\ \cmidrule{3-5}
            {Parameter} & {Unit} & {1} & {2} & {3}\\ \midrule
            $\omega_\mathrm{q,ge}^\mathrm{d}(0.0)/2\pi$ & [\si{\GHz}] &  5.487 &  5.903 &  6.364 \\
            $\omega_\mathrm{q,ge}^\mathrm{d}(0.5)/2\pi$ & [\si{\GHz}] &  3.757 &  4.161 &  4.764 \\
            $\alpha_\mathrm{q}^\mathrm{d}(0.0)/2\pi$    & [\si{\GHz}] & -0.163 & -0.150 & -0.154
            \\ \midrule
            $E_\mathrm{J,max}/h$                        & [\si{\GHz}] & 25.23  & 30.71  & 33.92 \\
            $E_\mathrm{c}/h$                            & [\si{\GHz}] &  0.160 &  0.151 &  0.159 \\
            \midrule
            && \multicolumn{3}{c}{{Right Module Qubits}} \\ \cmidrule{3-5}
            {Parameter} & {Unit} & {4} & {5} & {6} \\ \midrule
            $\omega_\mathrm{q,ge}^\mathrm{d}(0.0)/2\pi$ & [\si{\GHz}] &  5.644 &  5.914 &  6.450 \\
            $\omega_\mathrm{q,ge}^\mathrm{d}(0.5)/2\pi$ & [\si{\GHz}] &  4.088 &  4.301 &  4.884 \\
            $\alpha_\mathrm{q}^\mathrm{d}(0.0)/2\pi$    & [\si{\GHz}] & -0.171 & -0.161 & -0.158 \\ \midrule
            $E_\mathrm{J,max}/h$                        & [\si{\GHz}] & 25.63  & 29.20  & 34.64  \\
            $E_\mathrm{c}/h$                            & [\si{\GHz}] &   0.166 &  0.160 &  0.160\\
            \bottomrule
        \end{tabular}
    \end{table}

\section{Measurement apparatus}
    \label{sec:appendix:measurementapparatus}
        \begin{figure}
        \centering
        \includegraphics[width=85mm]{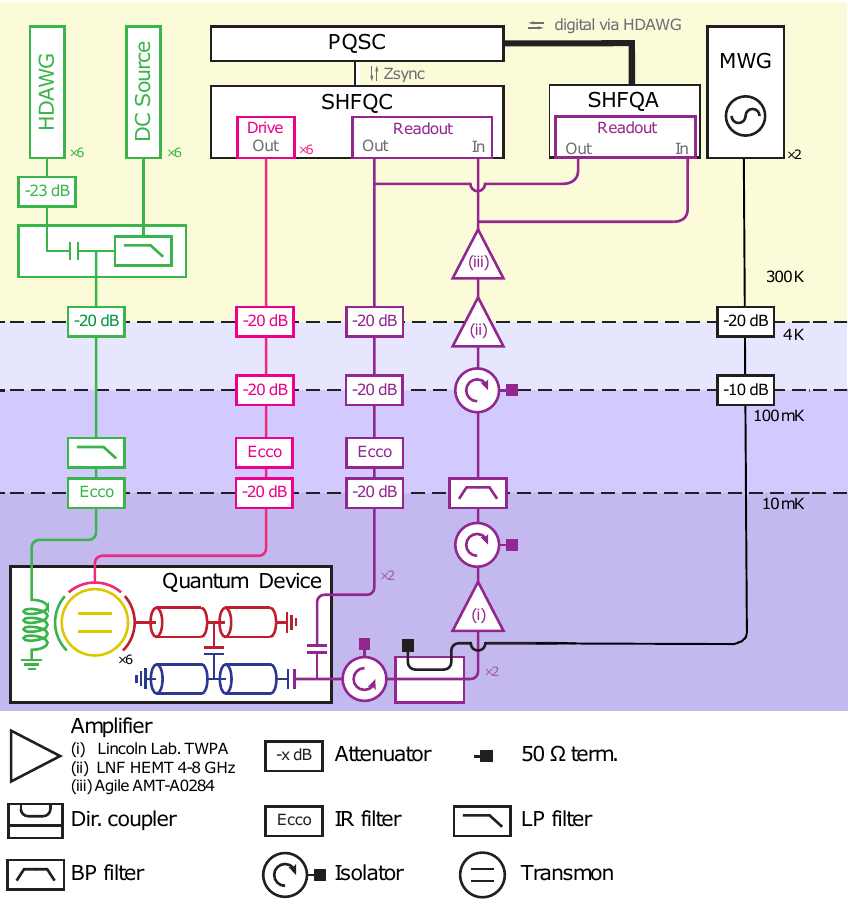}
        \caption{
            Measurement setup wiring diagram. See the text of Sec.~\ref{sec:appendix:measurementapparatus} for details. 
            This is exemplary for one qubit and one readout line, and does not display the whole setup.
        }
        \label{fig:appendix:wiring-diagram}
    \end{figure}

    The sample was measured at the \qty{10}{\milli K} stage of a dilution refrigerator, with the shielding configuration described in Ref.~\cite{Krinner2019}. 
    The wiring configuration and room-temperature control electronics are presented in App. Fig.~\ref{fig:appendix:wiring-diagram}.

    For qubit frequency control, we use an $RC$ bias-tee to combine a static voltage with nanosecond-timescale voltage pulses, determining an external flux through the qubit SQUID loop. 
    The static voltage is generated using a DC source, and the voltage pulses using an arbitrary waveform generator (AWG) with a sampling rate of \qty{2.0}{\giga S/s}. The voltage pulses are digitally pre-distorted to correct for the transfer function of the lines \cite{Hellings2025}.

    Qubit XY drive pulses are generated by a super-high-frequency qubit controller (SHFQC) using a double frequency conversion scheme \cite{Herrmann2022a}. 
    Qubit readout is carried out by the SHFQC and a super-high-frequency qubit analyzer (SHFQA), one for each frequency-multiplexed feedline serving each module of three qubits. 
    The readout pulses, generated with a superheterodyne signal generator, pass through the readout feedline, and are then amplified using a near-quantum limited traveling-wave parametric amplifier \cite{macklin2015}, a high-electron mobility amplifier (HEMT), and a room-temperature (RT) low-noise amplifier. 
    For the first feedline (read out using the SHFQC), this amplification chain results in a quantum measurement efficiency \cite{Bultink2018} of \num{0.33} for qubit 3.

    The SHFQC is triggered using a programmable quantum system controller (PQSC), while the SHFQA is triggered via the AWG.

\section{Numerical simulations}
    \label{sec:appendix:simulation}
        
    \begin{figure*}[ht]
    \centering
    \includegraphics[width=165mm]{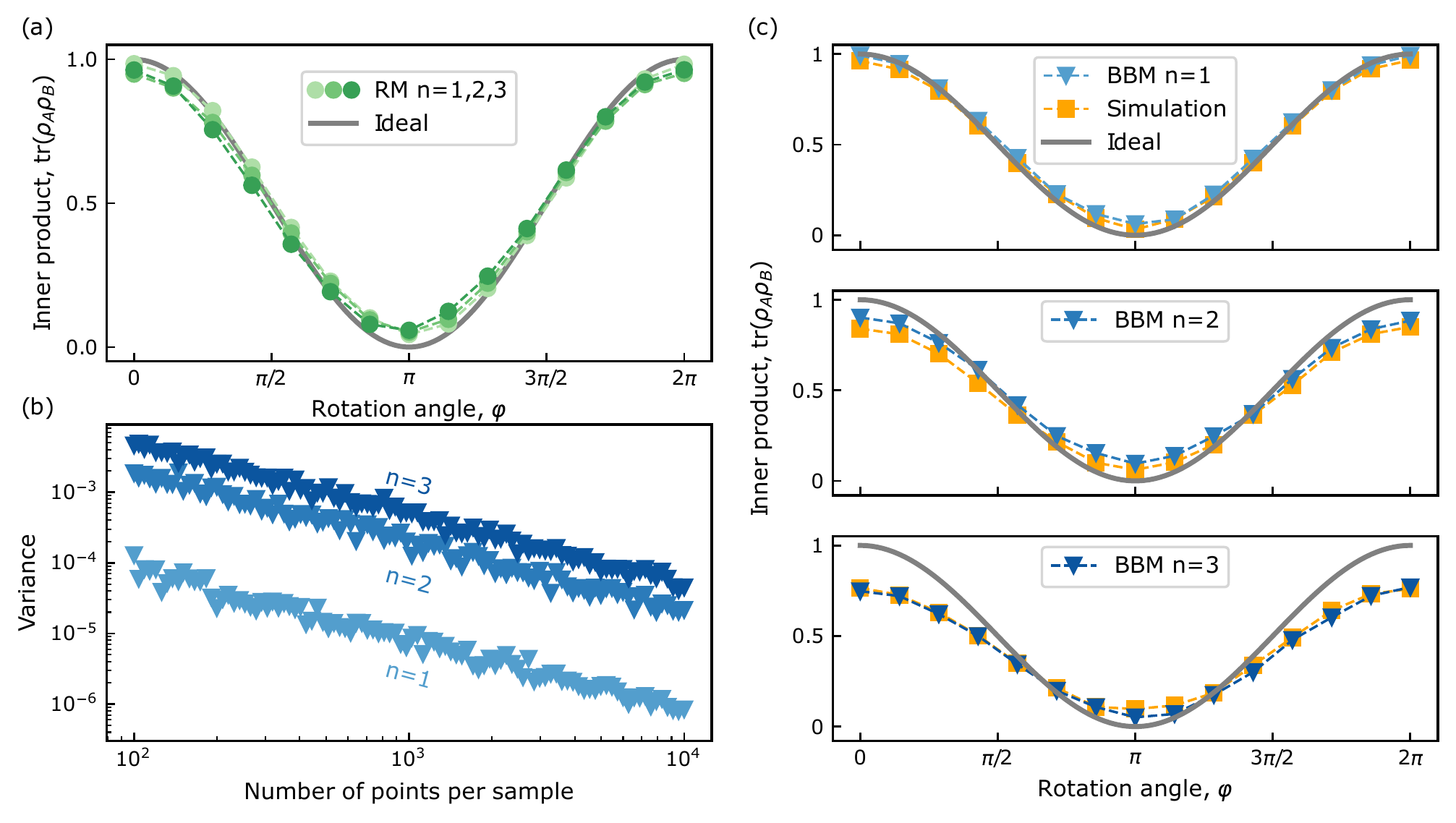}
    \caption{
        (a) Inner product between $\frac{1}{\sqrt{2}} \left( \ket{0}^{\otimes n} + \ket{1}^{\otimes n} \right)$ and $\frac{1}{\sqrt{2}} \left(\ket{0}^{\otimes n} + e^{i \varphi} \ket{1}^{\otimes n} \right)$, vs $\varphi$ for $n=1,2,3$ qubits, estimated using classical correlations between randomized measurements (see App. Sec.~\ref{sec:appendix:rm}). 
        (b) Calculated variance in the inner product for $n \in \{1,2,3\}$ qubits, estimated using bootstrapping from a \BBM{} dataset.
        (c) Simulations of the Bell-basis measurement circuit presented in Fig.~\ref{fig:main:circuit}, compared to the data in Fig.~\ref{fig:main:cpv}(c,d). 
        The error model is described in App.~\ref{sec:appendix:simulation}.
    }
    \label{fig:appendix:sim_and_rm}
    \end{figure*}

    We simulate the performance of our six-qubit device using Qiskit \cite{Javadi-Abhari2024}.
    A simple symmetric depolarizing channel is applied to the qubits involved in readout, single-qubit and two-qubit gates, using the median error rates discussed in Sec.~\ref{sec:results:device}, presented in Fig.~\ref{fig:main:performance}a.
    The channel is defined as \cite{Nielsen2010}
    \begin{equation}
       \rho \mapsto (1 - p) \rho + p \frac{I}{2^N}
    \end{equation}
    \noindent with the depolarizing parameter $p$. 

    We apply a thermal relaxation error channel to qubits which are not acted upon by each gate, to account for idling errors, defined as
    \begin{align}
    \rho_{00}(t) &= \rho_{00}(0)e^{-t/T_1} + p_{\text{th}}(1 - e^{-t/T_1})
    \\
    \rho_{11}(t) &= \rho_{11}(0)e^{-t/T_1} + (1 - p_{\text{th}})(1 - e^{-t/T_1})
    \\
    \rho_{10}(t) &= \rho_{10}(0)e^{-t/T_2}
    \\
    \rho_{01}(t) &= \rho_{01}(0)e^{-t/T_2}
    \end{align}
    \noindent where $T_1$ is the mean longitudinal relaxation time of the qubit, $T_2$ is the mean transverse relaxation time, $t$ is the gate time (\SI{40}{\ns} for single-qubit and \SI{100}{\ns} for two-qubit gates) and $p_\text{th}$ is the qubit residual excited state population. These values are measured in separate characterization experiments.

    We simulate the Bell-basis measurement protocol discussed in Sec.~\ref{sec:results:bbm}. 
    We include the effect of noise in both the state preparation and the \BBMprotocol{} illustrated in Fig.~\ref{fig:main:circuit}.
    
    We observe a cosinusoidal variation in simulated trace overlap, as expected for the states $\ghzn^{(0)}$ and $\ghzn^{(\varphi)}$. 
    Simulations involving more qubits and two-qubit gates show reduced contrast in the trace overlap (orange squares in  App. Fig.~\ref{fig:appendix:sim_and_rm}c). 
    These data are in quantitative agreement with the \BBM{} data (blue triangles).
    A validated simulation allows us to extract an error budget for our \BBMprotocol{}, including the errors which occur during single-qubit gates, two-qubit gates, and readout.

    To do this, we apply the method presented in Ref.~\cite{Chen2021p}. 
    We simulate the \BBMprotocol{} for two identical $\ghzn^{(0)}$ states, using the measured median error rates. 
    We then halve each error rate in turn (including halving the gate times, to reduce idling error), allowing the partial derivative of the total error rate with respect to each error source to be estimated
    \begin{equation}
        \frac{\partial E(p_i)}{\partial p_i} \approx \frac{E(p_i) - E\left(\frac{p_i}{2}\right)}{p_i - \frac{p_i}{2}}
    \end{equation}
    \noindent where $E = 1 - \tr(\rho_A \rho_B)$. 
    This partial derivative defines a weight. 
    One can then find the product of this weight with the measured median error rate of that error source, and normalize this by the weighted sum of all error rates
    \begin{equation}
        X_i = \frac{p_i\frac{\partial E(p_i)}{\partial p_i}}{\sum_ip_i\frac{\partial E(p_i)}{\partial p_i}}
    \end{equation}
    We see that readout error dominates the total error rate for $n=1$, while the two-qubit gate error contribution increases with $n$ and represents \SI{38}{\percent} of the error for $n=3$, see Table~\ref{tab:appendix:errorbudget}. 
    The dependence on readout error can be explained by the postprocessing procedure for \BBM{}, which uses the parity of the bit-string resulting from a bit-wise AND operation between the measurement results on each module. 
    A single bit-flip due to a readout error leads to a flip of the parity. 
    When these parity values are averaged for the trace overlap, these individual bit-flips contribute significantly to the error.

\begin{table}[b]
    \centering
    \caption{
        Error budget for \BBM{}, with the proportion of total errors that come from readout, single-qubit gates, and two-qubit gates. 
        For experiments involving $\ghzn^{(0)}$ states with $n\in \{1,2,3\}$ qubits. 
        The error in the calculated percentages is less than \qty{0.2}{}.
    }
    \label{tab:appendix:errorbudget}
    \begin{tabular}{lS[table-format=2.2]S[table-format=2.2]S[table-format=2]S[table-format=2.2]}
    \toprule
    && \multicolumn{3}{c}{Qubits per module, $n$} \\
    Error Type & {Unit} & {One} & {Two} & {Three} \\
    \midrule
    Measurement & \%  & 56.43 & 52.44 & 48.23 \\
    Single-Qubit Gates & \%  & 19.14 & 13.85 & 13.46 \\
    Two-Qubit Gates & \%  & 24.43 & 33.71 & 38.31 \\
    \bottomrule
    \end{tabular}
    
\end{table}

\section{Randomized measurements}
\label{sec:appendix:rm}

    As described in Sec.~\ref{sec:results:dipe}, we estimate the inner product between a $n$-qubit $\ghzn^{(0)}$ state on one module and a $n$-qubit $\ghzn^{(\varphi)}$ state on another module using quantum state tomography (QST). 
    Using the theoretical frameworks of classical shadows~\cite{Aaronson2018} and randomized measurements (RM)~\cite{Elben2019,Elben2022}, the experimental resources required for cross-platform verification can be reduced compared to those required for full \QST{}.
    We demonstrate this with two identically-prepared $\ghzn^{(0)}$ states in Fig.~\ref{fig:main:cpv}e (Sec.~\ref{sec:results:complexity}).

    We discuss the protocol introduced in Ref.~\cite{Elben2019}, in which the overlap $\tr\left(\rho_{A} \rho_{B}\right)$ is estimated from the second-order cross-correlations between measurements of the $n$-qubit states $\rho_{A}$ and $\rho_{B}$. 
    After state preparation, the same random unitary $U$ is applied to both states, and they are measured in the computational basis, resulting in two bitstrings of length $n$, $s$ and $s^{\prime}$. 
    Repeating this for many unitaries, the inner product is estimated using the equation
    \begin{equation}
    \label{eq:rm_trace_overlap}
    \tr\left(\rho_A \rho_B\right)=2^n \sum_{s, s^{\prime}}(-2)^{-D\left[s, s^{\prime}\right]} \overline{P_U^{(A)}(s) P_U^{(B)}\left(s^{\prime}\right)},
    \end{equation}
    where $D\left[s, s^{\prime}\right]$ is the Hamming distance between $s$ and $s^{\prime}$, $P_U^{(i)}(s)=\tr\left(U \rho_i U^{\dagger}|s\rangle\langle s|\right)$, and the overline denotes the average over the chosen unitaries $U$, as in Ref.~\cite{Zhu2022e}.
    
    This approach scales more favorably than \QST{}, with a required number of experimental repetitions $N \sim 2^{b n}$ with $b\lesssim1$, rather than $b\ge2$ as for \QST{} \cite{Elben2022}. 
    The exact value of $b$ depends on the states compared and unitaries chosen. 
    Typically, scaling is improved using unitaries selected with a greedy algorithm \cite{Zhu2022e}, which selects the next unitary to minimize a cost function based on the trace distance between the new unitary and the previously selected unitaries.
    
    We analyze the phase-sweep data presented in Fig.~\ref{fig:main:cpv}(a,b) using randomized measurements. 
    From the bases used in the \QST{} measurements, we select $3^n$ positive measurement bases using a greedy algorithm \cite{Zhu2022e}. 
    We implement this algorithm by choosing the first unitary $U_1$ randomly, then choose each subsequent unitary by minimizing the cost function
    \begin{equation}
    \mathcal{C} \left( \mathcal{B}, U_i \right) = - \sum_{U \in \mathcal{B}} d\left( U, U_{i} \right),
    \end{equation}
    \noindent where $\mathcal{B}$ is the set of previously-chosen unitaries and $d(U,U_{i})$ is the distance between two unitaries, defined as
    \begin{equation}
    d \left( U, U_{i} \right) = \sqrt{2 - |\tr(U^\dagger U_{i})|}.
    \end{equation}
    We choose $3^n$ as a balance between sample efficiency and the variance of the resulting inner product. 
    We estimate the inner product using Eq.~\eqref{eq:rm_trace_overlap}. 

    The estimated inner product is close to one for nominally-identical states with zero phase, and close to zero for nominally-orthogonal states, as expected. 
    The resulting data is presented in App. Fig.~\ref{fig:appendix:sim_and_rm}. 
    These data agree well with the \QST{} data in Fig.~\ref{fig:main:cpv}(a,b).
        
\section{Unitary fidelity estimation}
\label{sec:appendix:unitaries}

\begin{figure*}
        \centering
        \includegraphics[width=160mm]{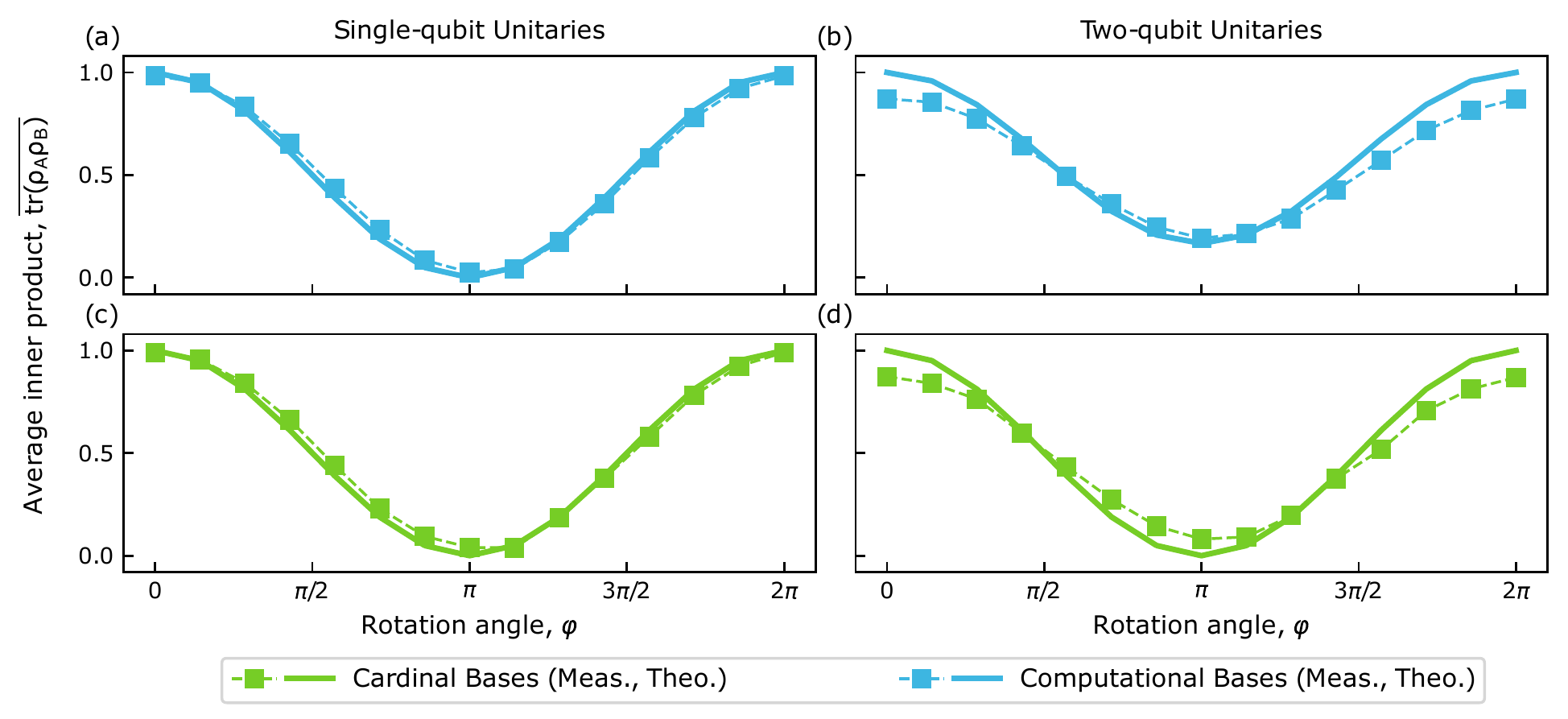}
        \caption{
            Unitary fidelity estimated using Bell-basis measurements. 
            The unitaries to prepare the states $\frac{1}{\sqrt{2}} \left( \ket{0}^{\otimes n} + \ket{1}^{\otimes n} \right)$ and $\frac{1}{\sqrt{2}} \left(\ket{0}^{\otimes n} + e^{i \varphi} \ket{1}^{\otimes n} \right)$ are applied to the left and right modules respectively, starting from all possible combinations of the cardinal states $\left\{ \ket{0}, \ket{1}, \ket{+}, \ket{-}, \ket{i}, \ket{-i} \right\}$. 
            The inner product for each initial state is calculated, and averaged over the cardinal states (blue curves) and computational states (green curves), for one (a, c) and two (b, d) qubits per module.
        }
        \label{fig:appendix:unitaries}
    \end{figure*}

In Sec.~\ref{sec:results:bbm}, we discussed how Bell-basis measurements (BBM) can be used to estimate the inner product between two quantum states.
Given the purities of the states, this allows the fidelity between them to be determined via Eq.~\eqref{eq:def-fidelity}. 
This can be straightforwardly extended to estimate process fidelities by averaging over the outcomes for different initial states \cite{Knoerzer2023}, where process fidelity is defined as \cite{Raginsky2001}
\begin{equation}
\mathcal{F}_p=\frac{\operatorname{tr}\left[U_A^{\dagger} U_B\right]}{d^2},
\end{equation}
\noindent where $U_A \left(U_B\right)$ is the unitary operation performed on module $\mathrm{A} (\mathrm{B})$.

To faithfully estimate the process fidelity between two unitary quantum operations $U_A$ and $U_B$, one may evaluate the fidelity over an ensemble of input states.
A common and effective method is to use states generated by a unitary 2-design~\cite{Dankert2009,Gross2007}, which approximates the statistical properties of the Haar measure.
In practice, such a 2-design can be implemented by applying local random unitaries from the Clifford group to randomly selected pairs of qubits~\cite{Brandao2016c}.

Specifically, we choose $K$ unitaries $\left\{U_k\right\}_{k=1}^K$ that form a unitary 2-design.
These unitaries generate $K$ initial states $|k\rangle=U_k|0\rangle$ from the ground state $\ket{0}$. 
The average fidelity is then computed as
\begin{equation}
\mathcal{F}_{a v}\left(U_A, U_B\right)=\frac{1}{K} \sum_{k=1}^K \left| \bra{k} U_A^{\dagger} U_B \ket{k} \right| ^ 2.
\end{equation}

This average fidelity relates to the process fidelity $\mathcal{F}_p$ as \cite{Horodecki1999,Nielsen2002}
\begin{equation}
\mathcal{F}_p\left(U_A, U_B\right)=\frac{(d+1) \mathcal{F}_{a v}\left(U_A, U_B\right)-1}{d}
\end{equation}
\noindent for a Hilbert space of dimension $d$.

For one-qubit unitaries, the state-preparation unitaries for the cardinal states $\{ \ket{0}, \ket{1}, \ket{+}, \ket{-}, \ket{i}, \ket{-i} \}$, from the ground state $\ket{0}$, form a 2-design. 
For higher-dimensional unitaries, preparing the necessary initial states requires a significant number of two-qubit gates, and would hence dominate the error of our \BBMprotocol{}. 
Thus, as an alternative method that does not yield the process fidelity but a metric that can be more realistically-obtained on current hardware, we investigate computational basis sampling \cite{Knoerzer2023}.
Averaging over outcomes using computational basis states yields an estimate of
\begin{equation}
\mathcal{F}_{s q}=\frac{1}{d} \sum_{p \in\{0,1\}^n} \left|\bra{p} U_A^{\dagger} U_B \ket{p} \right| ^ 2.
\end{equation}
While this measure is basis-dependent, in contrast to the process fidelity, we can efficiently obtain it in our experiment.
We thus estimate it for the $\ghzn^{(\varphi)}$ preparation unitary presented in Fig.~\ref{fig:main:circuit}, for one- and two-qubit states. 
Before applying the state preparation circuit, we prepare a product state of single-qubit computational or cardinal states on each module, with the same single-qubit state on the corresponding qubits of each module. 
We note that, for two-qubit unitaries, parallel single-qubit Clifford circuits without entangling gates do not form a 2-design, and hence the average inner product we estimate is not related to a true process fidelity.

After applying the \BBM{} circuit and estimating the inner product between the final states (using $10^5$ repetitions), we average over all permutations of initial states, resulting in an average inner product. 
As before, we apply an additional phase $\varphi$ during the application of one of the unitaries, which we vary over an entire period. 

As before, we observe a cosinusoidal variation in average inner product with the phase on one state, with reduced maximal overlap for two-qubit unitaries compared to single-qubit unitaries. 
We additionally observe a rightward shift in the curves, which we attribute to spectator errors \cite{Krinner2020}. These are more prevalent in the case of process fidelities, since spectator qubits are more likely to be in the excited state.
We present this in App. Fig.~\ref{fig:appendix:unitaries}, alongside the expected theory curve calculated using noiseless simulation. 
Note that we present the average inner product, which would need to be normalized by the state purity to estimate process fidelity. 

\bibliographystyle{apsrev4-2-title-etal-template}
\bibliography{main}

\begin{thebibliography}{97}%
\makeatletter
\providecommand \@ifxundefined [1]{%
 \@ifx{#1\undefined}
}%
\providecommand \@ifnum [1]{%
 \ifnum #1\expandafter \@firstoftwo
 \else \expandafter \@secondoftwo
 \fi
}%
\providecommand \@ifx [1]{%
 \ifx #1\expandafter \@firstoftwo
 \else \expandafter \@secondoftwo
 \fi
}%
\providecommand \natexlab [1]{#1}%
\providecommand \enquote  [1]{``#1''}%
\providecommand \bibnamefont  [1]{#1}%
\providecommand \bibfnamefont [1]{#1}%
\providecommand \citenamefont [1]{#1}%
\providecommand \href@noop [0]{\@secondoftwo}%
\providecommand \href [0]{\begingroup \@sanitize@url \@href}%
\providecommand \@href[1]{\@@startlink{#1}\@@href}%
\providecommand \@@href[1]{\endgroup#1\@@endlink}%
\providecommand \@sanitize@url [0]{\catcode `\\12\catcode `\$12\catcode `\&12\catcode `\#12\catcode `\^12\catcode `\_12\catcode `\%12\relax}%
\providecommand \@@startlink[1]{}%
\providecommand \@@endlink[0]{}%
\providecommand \url  [0]{\begingroup\@sanitize@url \@url }%
\providecommand \@url [1]{\endgroup\@href {#1}{\urlprefix }}%
\providecommand \urlprefix  [0]{URL }%
\providecommand \Eprint [0]{\href }%
\providecommand \doibase [0]{https://doi.org/}%
\providecommand \selectlanguage [0]{\@gobble}%
\providecommand \bibinfo  [0]{\@secondoftwo}%
\providecommand \bibfield  [0]{\@secondoftwo}%
\providecommand \translation [1]{[#1]}%
\providecommand \BibitemOpen [0]{}%
\providecommand \bibitemStop [0]{}%
\providecommand \bibitemNoStop [0]{.\EOS\space}%
\providecommand \EOS [0]{\spacefactor3000\relax}%
\providecommand \BibitemShut  [1]{\csname bibitem#1\endcsname}%
\let\auto@bib@innerbib\@empty
\bibitem [{\citenamefont {Watrous}(2008)Watrous, J.}]{Watrous2008}%
  \BibitemOpen
  \bibfield  {author} {\bibinfo {author} {Watrous, J.},\ }\bibfield  {title} {\bibinfo {title} {Quantum computational complexity},\ }\href {https://arxiv.org/abs/0804.3401} {\bibfield  {journal} {\bibinfo  {journal} {arXiv:0804.3401}\ } (\bibinfo {year} {2008})}\BibitemShut {NoStop}%
\bibitem [{\citenamefont {Aaronson}(2013)Aaronson, Scott}]{Aaronson2013b}%
  \BibitemOpen
  \bibfield  {author} {\bibinfo {author} {Aaronson, S.},\ }\href@noop {} {\emph {\bibinfo {title} {Quantum computing since Democritus}}}\ (\bibinfo  {publisher} {Cambridge University Press},\ \bibinfo {year} {2013})\BibitemShut {NoStop}%
\bibitem [{\citenamefont {Nielsen}\ and\ \citenamefont {Chuang}(2010)Nielsen, Michael~A. and Chuang, Isaac~L.}]{Nielsen2010}%
  \BibitemOpen
  \bibfield  {author} {\bibinfo {author} {Nielsen, M.~A.}\ and\ \bibinfo {author} {Chuang, I.~L.},\ }\href@noop {} {\emph {\bibinfo {title} {Quantum Computation and Quantum Information}}},\ \bibinfo {edition} {{10th anniversary}}\ ed.\ (\bibinfo  {publisher} {Cambridge University Press},\ \bibinfo {address} {New York, USA},\ \bibinfo {year} {2010})\BibitemShut {NoStop}%
\bibitem [{\citenamefont {Dalton}\ \emph {et~al.}(2024)Dalton, Kieran and Long, Christopher K. and Yordanov, Yordan S. and Smith, Charles G. and Barnes, Crispin H. W. and Mertig, Normann and Arvidsson-Shukur, David R. M.}]{Dalton2024}%
  \BibitemOpen
  \bibfield  {author} {\bibinfo {author} {Dalton, K.}, \bibinfo {author} {Long, C.~K.}, \bibinfo {author} {Yordanov, Y.~S.}, \emph {et~al.},\ }\bibfield  {title} {\bibinfo {title} {Quantifying the effect of gate errors on variational quantum eigensolvers for quantum chemistry},\ }\href {https://doi.org/10.1038/s41534-024-00808-x} {\bibfield  {journal} {\bibinfo  {journal} {npj Quantum Information}\ }\textbf {\bibinfo {volume} {10}},\ \bibinfo {pages} {18} (\bibinfo {year} {2024})}\BibitemShut {NoStop}%
\bibitem [{\citenamefont {Gidney}(2025)Gidney, C.}]{Gidney2025b}%
  \BibitemOpen
  \bibfield  {author} {\bibinfo {author} {Gidney, C.},\ }\bibfield  {title} {\bibinfo {title} {How to factor 2048 bit rsa integers with less than a million noisy qubits},\ }\href {https://arxiv.org/abs/2505.15917} {\bibfield  {journal} {\bibinfo  {journal} {arXiv:2505.15917}\ } (\bibinfo {year} {2025})}\BibitemShut {NoStop}%
\bibitem [{\citenamefont {Dalzell}\ \emph {et~al.}(2025)Dalzell, A. M. and McArdle, S. and Berta, M. and Bienias, P. and Chen, C. and Gily\'en, A. and Hann, C. T. and Kastoryano, M. J. and Khabiboulline, E. T. and Kubica, A. and Salton, G. and Wang, S. and Brand\~ao, F. G. S. L.}]{Dalzell2025}%
  \BibitemOpen
  \bibfield  {author} {\bibinfo {author} {Dalzell, A.~M.}, \bibinfo {author} {McArdle, S.}, \bibinfo {author} {Berta, M.}, \emph {et~al.},\ }\href@noop {} {\emph {\bibinfo {title} {Quantum algorithms: A survey of applications and end-to-end complexities}}}\ (\bibinfo  {publisher} {Cambridge University Press},\ \bibinfo {year} {2025})\BibitemShut {NoStop}%
\bibitem [{\citenamefont {Preskill}(2025)Preskill, J.}]{Preskill2025}%
  \BibitemOpen
  \bibfield  {author} {\bibinfo {author} {Preskill, J.},\ }\bibfield  {title} {\bibinfo {title} {Beyond nisq: The megaquop machine},\ }\href {https://arxiv.org/abs/2502.17368} {\bibfield  {journal} {\bibinfo  {journal} {arXiv:2502.17368}\ } (\bibinfo {year} {2025})}\BibitemShut {NoStop}%
\bibitem [{\citenamefont {Awschalom}\ \emph {et~al.}(2021)Awschalom, David and Berggren, Karl K. and Bernien, Hannes and Bhave, Sunil and Carr, Lincoln D. and Davids, Paul and Economou, Sophia E. and Englund, Dirk and Faraon, Andrei and Fejer, Martin and Guha, Saikat and Gustafsson, Martin V. and Hu, Evelyn and Jiang, Liang and Kim, Jungsang and Korzh, Boris and Kumar, Prem and Kwiat, Paul G. and Lon\ifmmode \check{c}\else \v{c}\fi{}ar, Marko and Lukin, Mikhail D. and Miller, David A.B. and Monroe, Christopher and Nam, Sae Woo and Narang, Prineha and Orcutt, Jason S. and Raymer, Michael G. and Safavi-Naeini, Amir H. and Spiropulu, Maria and Srinivasan, Kartik and Sun, Shuo and Vu\ifmmode \check{c}\else \v{c}\fi{}kovi\ifmmode \acute{c}\else \'{c}\fi{}, Jelena and Waks, Edo and Walsworth, Ronald and Weiner, Andrew M. and Zhang, Zheshen}]{Awschalom2021}%
  \BibitemOpen
  \bibfield  {author} {\bibinfo {author} {Awschalom, D.}, \bibinfo {author} {Berggren, K.~K.}, \bibinfo {author} {Bernien, H.}, \emph {et~al.},\ }\bibfield  {title} {\bibinfo {title} {Development of quantum interconnects (quics) for next-generation information technologies},\ }\href {https://doi.org/10.1103/PRXQuantum.2.017002} {\bibfield  {journal} {\bibinfo  {journal} {PRX Quantum}\ }\textbf {\bibinfo {volume} {2}},\ \bibinfo {pages} {017002} (\bibinfo {year} {2021})}\BibitemShut {NoStop}%
\bibitem [{\citenamefont {Gold}\ \emph {et~al.}(2021)Gold, Alysson and Paquette, J. P. and Stockklauser, Anna and Reagor, Matthew J. and Alam, M. Sohaib and Bestwick, Andrew and Didier, Nicolas and Nersisyan, Ani and Oruc, Feyza and Razavi, Armin and Scharmann, Ben and Sete, Eyob A. and Sur, Biswajit and Venturelli, Davide and Winkleblack, Cody James and Wudarski, Filip and Harburn, Mike and Rigetti, Chad}]{Gold2021}%
  \BibitemOpen
  \bibfield  {author} {\bibinfo {author} {Gold, A.}, \bibinfo {author} {Paquette, J.~P.}, \bibinfo {author} {Stockklauser, A.}, \emph {et~al.},\ }\bibfield  {title} {\bibinfo {title} {Entanglement across separate silicon dies in a modular superconducting qubit device},\ }\href {https://doi.org/10.1038/s41534-021-00484-1} {\bibfield  {journal} {\bibinfo  {journal} {npj Quantum Information}\ }\textbf {\bibinfo {volume} {7}},\ \bibinfo {pages} {142} (\bibinfo {year} {2021})}\BibitemShut {NoStop}%
\bibitem [{\citenamefont {Field}\ \emph {et~al.}(2024)Field, Mark and Chen, Angela Q. and Scharmann, Ben and Sete, Eyob A. and Oruc, Feyza and Vu, Kim and Kosenko, Valentin and Mutus, Joshua Y. and Poletto, Stefano and Bestwick, Andrew}]{Field2024}%
  \BibitemOpen
  \bibfield  {author} {\bibinfo {author} {Field, M.}, \bibinfo {author} {Chen, A.~Q.}, \bibinfo {author} {Scharmann, B.}, \emph {et~al.},\ }\bibfield  {title} {\bibinfo {title} {Modular superconducting-qubit architecture with a multichip tunable coupler},\ }\href {https://doi.org/10.1103/physrevapplied.21.054063} {\bibfield  {journal} {\bibinfo  {journal} {Physical Review Applied}\ }\textbf {\bibinfo {volume} {21}},\ \bibinfo {pages} {054063} (\bibinfo {year} {2024})}\BibitemShut {NoStop}%
\bibitem [{\citenamefont {Wu}\ \emph {et~al.}(2024)Wu, Xuntao and Yan, Haoxiong and Andersson, Gustav and Anferov, Alexander and Chou, Ming-Han and Conner, Christopher R. and Grebel, Joel and Joshi, Yash J. and Li, Shiheng and Miller, Jacob M. and Povey, Rhys G. and Qiao, Hong and Cleland, Andrew N.}]{Wu2024e}%
  \BibitemOpen
  \bibfield  {author} {\bibinfo {author} {Wu, X.}, \bibinfo {author} {Yan, H.}, \bibinfo {author} {Andersson, G.}, \emph {et~al.},\ }\bibfield  {title} {\bibinfo {title} {Modular quantum processor with an all-to-all reconfigurable router},\ }\href {https://doi.org/10.1103/PhysRevX.14.041030} {\bibfield  {journal} {\bibinfo  {journal} {Phys. Rev. X}\ }\textbf {\bibinfo {volume} {14}},\ \bibinfo {pages} {041030} (\bibinfo {year} {2024})}\BibitemShut {NoStop}%
\bibitem [{\citenamefont {Burkhart}\ \emph {et~al.}(2021)Burkhart, Luke D. and Teoh, James D. and Zhang, Yaxing and Axline, Christopher J. and Frunzio, Luigi and Devoret, M.H. and Jiang, Liang and Girvin, S.M. and Schoelkopf, R.J.}]{Burkhart2021}%
  \BibitemOpen
  \bibfield  {author} {\bibinfo {author} {Burkhart, L.~D.}, \bibinfo {author} {Teoh, J.~D.}, \bibinfo {author} {Zhang, Y.}, \emph {et~al.},\ }\bibfield  {title} {\bibinfo {title} {Error-detected state transfer and entanglement in a superconducting quantum network},\ }\href {https://doi.org/10.1103/PRXQuantum.2.030321} {\bibfield  {journal} {\bibinfo  {journal} {PRX Quantum}\ }\textbf {\bibinfo {volume} {2}},\ \bibinfo {pages} {030321} (\bibinfo {year} {2021})}\BibitemShut {NoStop}%
\bibitem [{\citenamefont {Niu}\ \emph {et~al.}(2023)Niu, Jingjing and Zhang, Libo and Liu, Yang and Qiu, Jiawei and Huang, Wenhui and Huang, Jiaxiang and Jia, Hao and Liu, Jiawei and Tao, Ziyu and Wei, Weiwei and et al.}]{Niu2023}%
  \BibitemOpen
  \bibfield  {author} {\bibinfo {author} {Niu, J.}, \bibinfo {author} {Zhang, L.}, \bibinfo {author} {Liu, Y.}, \emph {et~al.},\ }\bibfield  {title} {\bibinfo {title} {Low-loss interconnects for modular superconducting quantum processors},\ }\href {https://doi.org/10.1038/s41928-023-00925-z} {\bibfield  {journal} {\bibinfo  {journal} {Nature Electronics}\ }\textbf {\bibinfo {volume} {6}},\ \bibinfo {pages} {235–241} (\bibinfo {year} {2023})}\BibitemShut {NoStop}%
\bibitem [{\citenamefont {Magnard}\ \emph {et~al.}(2020)Magnard, P. and Storz, S. and Kurpiers, P. and Sch\"ar, J. and Marxer, F. and L\"utolf, J. and Walter, T. and Besse, J.-C. and Gabureac, M. and Reuer, K. and Akin, A. and Royer, B. and Blais, A. and Wallraff, A.}]{Magnard2020}%
  \BibitemOpen
  \bibfield  {author} {\bibinfo {author} {Magnard, P.}, \bibinfo {author} {Storz, S.}, \bibinfo {author} {Kurpiers, P.}, \emph {et~al.},\ }\bibfield  {title} {\bibinfo {title} {Microwave quantum link between superconducting circuits housed in spatially separated cryogenic systems},\ }\href {https://doi.org/10.1103/PhysRevLett.125.260502} {\bibfield  {journal} {\bibinfo  {journal} {Phys. Rev. Lett.}\ }\textbf {\bibinfo {volume} {125}},\ \bibinfo {pages} {260502} (\bibinfo {year} {2020})}\BibitemShut {NoStop}%
\bibitem [{\citenamefont {Storz}\ \emph {et~al.}(2023)Storz, Simon and Sch{\"a}r, Josua and Kulikov, Anatoly and Magnard, Paul and Kurpiers, Philipp and L{\"u}tolf, Janis and Walter, Theo and Copetudo, Adrian and Reuer, Kevin and Akin, Abdulkadir and Besse, Jean-Claude and Gabureac, Mihai and Norris, Graham J. and Rosario, Andr{\'e}s and Martin, Ferran and Martinez, Jos{\'e} and Amaya, Waldimar and Mitchell, Morgan W. and Abellan, Carlos and Bancal, Jean-Daniel and Sangouard, Nicolas and Royer, Baptiste and Blais, Alexandre and Wallraff, Andreas}]{Storz2023}%
  \BibitemOpen
  \bibfield  {author} {\bibinfo {author} {Storz, S.}, \bibinfo {author} {Sch{\"a}r, J.}, \bibinfo {author} {Kulikov, A.}, \emph {et~al.},\ }\bibfield  {title} {\bibinfo {title} {Loophole-free bell inequality violation with superconducting circuits},\ }\href {https://doi.org/10.1038/s41586-023-05885-0} {\bibfield  {journal} {\bibinfo  {journal} {Nature}\ }\textbf {\bibinfo {volume} {617}},\ \bibinfo {pages} {265} (\bibinfo {year} {2023})}\BibitemShut {NoStop}%
\bibitem [{\citenamefont {Andrews}\ \emph {et~al.}(2014)Andrews, R. W. and Peterson, R. W. and Purdy, T. P. and Cicak, K. and Simmonds, R. W. and Regal, C. A. and Lehnert, K. W.}]{Andrews2014}%
  \BibitemOpen
  \bibfield  {author} {\bibinfo {author} {Andrews, R.~W.}, \bibinfo {author} {Peterson, R.~W.}, \bibinfo {author} {Purdy, T.~P.}, \emph {et~al.},\ }\bibfield  {title} {\bibinfo {title} {Bidirectional and efficient conversion between microwave and optical light},\ }\href {https://doi.org/10.1038/nphys2911} {\bibfield  {journal} {\bibinfo  {journal} {Nat. Phys.}\ }\textbf {\bibinfo {volume} {10}},\ \bibinfo {pages} {321} (\bibinfo {year} {2014})}\BibitemShut {NoStop}%
\bibitem [{\citenamefont {Mirhosseini}\ \emph {et~al.}(2020)Mirhosseini, Mohammad and Sipahigil, Alp and Kalaee, Mahmoud and Painter, Oskar}]{Mirhosseini2020}%
  \BibitemOpen
  \bibfield  {author} {\bibinfo {author} {Mirhosseini, M.}, \bibinfo {author} {Sipahigil, A.}, \bibinfo {author} {Kalaee, M.},\ and\ \bibinfo {author} {Painter, O.},\ }\bibfield  {title} {\bibinfo {title} {Superconducting qubit to optical photon transduction},\ }\href {https://doi.org/10.1038/s41586-020-3038-6} {\bibfield  {journal} {\bibinfo  {journal} {Nature}\ }\textbf {\bibinfo {volume} {588}},\ \bibinfo {pages} {599} (\bibinfo {year} {2020})}\BibitemShut {NoStop}%
\bibitem [{\citenamefont {Han}\ \emph {et~al.}(2021)Xu Han and Wei Fu and Chang-Ling Zou and Liang Jiang and Hong X. Tang}]{Han2021f}%
  \BibitemOpen
  \bibfield  {author} {\bibinfo {author} {Han, X.}, \bibinfo {author} {Fu, W.}, \bibinfo {author} {Zou, C.-L.}, \emph {et~al.},\ }\bibfield  {title} {\bibinfo {title} {Microwave-optical quantum frequency conversion},\ }\href {https://doi.org/10.1364/OPTICA.425414} {\bibfield  {journal} {\bibinfo  {journal} {Optica}\ }\textbf {\bibinfo {volume} {8}},\ \bibinfo {pages} {1050} (\bibinfo {year} {2021})}\BibitemShut {NoStop}%
\bibitem [{\citenamefont {Sahu}\ \emph {et~al.}(2022)Sahu, R. and Hease, W. and Rueda, A. and Arnold, G. and Qiu, L. and Fink, J.}]{Sahu2022}%
  \BibitemOpen
  \bibfield  {author} {\bibinfo {author} {Sahu, R.}, \bibinfo {author} {Hease, W.}, \bibinfo {author} {Rueda, A.}, \emph {et~al.},\ }\bibfield  {title} {\bibinfo {title} {Quantum-enabled operation of a microwave-optical interface},\ }\bibfield  {journal} {\bibinfo  {journal} {Nature Communications}\ }\textbf {\bibinfo {volume} {13}},\ \href {https://doi.org/10.1038/s41467-022-28924-2} {10.1038/s41467-022-28924-2} (\bibinfo {year} {2022})\BibitemShut {NoStop}%
\bibitem [{\citenamefont {Ang}\ \emph {et~al.}(2024)Ang, James and Carini, Gabriella and Chen, Yanzhu and Chuang, Isaac and Demarco, Michael and Economou, Sophia and Eickbusch, Alec and Faraon, Andrei and Fu, Kai-Mei and Girvin, Steven and Hatridge, Michael and Houck, Andrew and Hilaire, Paul and Krsulich, Kevin and Li, Ang and Liu, Chenxu and Liu, Yuan and Martonosi, Margaret and McKay, David and Misewich, Jim and Ritter, Mark and Schoelkopf, Robert and Stein, Samuel and Sussman, Sara and Tang, Hong and Tang, Wei and Tomesh, Teague and Tubman, Norm and Wang, Chen and Wiebe, Nathan and Yao, Yongxin and Yost, Dillon and Zhou, Yiyu}]{Ang2024}%
  \BibitemOpen
  \bibfield  {author} {\bibinfo {author} {Ang, J.}, \bibinfo {author} {Carini, G.}, \bibinfo {author} {Chen, Y.}, \emph {et~al.},\ }\bibfield  {title} {\bibinfo {title} {Arquin: Architectures for multinode superconducting quantum computers},\ }\bibfield  {journal} {\bibinfo  {journal} {ACM Transactions on Quantum Computing}\ }\textbf {\bibinfo {volume} {5}},\ \href {https://doi.org/10.1145/3674151} {10.1145/3674151} (\bibinfo {year} {2024})\BibitemShut {NoStop}%
\bibitem [{\citenamefont {Magesan}\ \emph {et~al.}(2012{\natexlab{a}})Magesan, Easwar and Gambetta, Jay M. and Johnson, B. R. and Ryan, Colm A. and Chow, Jerry M. and Merkel, Seth T. and da Silva, Marcus P. and Keefe, George A. and Rothwell, Mary B. and Ohki, Thomas A. and Ketchen, Mark B. and Steffen, M.}]{Magesan2012}%
  \BibitemOpen
  \bibfield  {author} {\bibinfo {author} {Magesan, E.}, \bibinfo {author} {Gambetta, J.~M.}, \bibinfo {author} {Johnson, B.~R.}, \emph {et~al.},\ }\bibfield  {title} {\bibinfo {title} {Efficient measurement of quantum gate error by interleaved randomized benchmarking},\ }\href {https://doi.org/10.1103/PhysRevLett.109.080505} {\bibfield  {journal} {\bibinfo  {journal} {Phys. Rev. Lett.}\ }\textbf {\bibinfo {volume} {109}},\ \bibinfo {pages} {080505} (\bibinfo {year} {2012}{\natexlab{a}})}\BibitemShut {NoStop}%
\bibitem [{\citenamefont {Magesan}\ \emph {et~al.}(2012{\natexlab{b}})Magesan, Easwar and Gambetta, Jay M. and Emerson, Joseph}]{Magesan2012a}%
  \BibitemOpen
  \bibfield  {author} {\bibinfo {author} {Magesan, E.}, \bibinfo {author} {Gambetta, J.~M.},\ and\ \bibinfo {author} {Emerson, J.},\ }\bibfield  {title} {\bibinfo {title} {Characterizing quantum gates via randomized benchmarking},\ }\href {https://doi.org/10.1103/PhysRevA.85.042311} {\bibfield  {journal} {\bibinfo  {journal} {Phys. Rev. A}\ }\textbf {\bibinfo {volume} {85}},\ \bibinfo {pages} {042311} (\bibinfo {year} {2012}{\natexlab{b}})}\BibitemShut {NoStop}%
\bibitem [{\citenamefont {Heya}\ \emph {et~al.}(2025)Heya, K. and Phung, T. and Malekakhlagh, M. and Steiner, R. and Turchetti, M. and Shanks, W. and Mamin, J. and Lu, W. and Kandel, Y. P. and Sundaresan, N. and Orcutt, J.}]{Heya2025}%
  \BibitemOpen
  \bibfield  {author} {\bibinfo {author} {Heya, K.}, \bibinfo {author} {Phung, T.}, \bibinfo {author} {Malekakhlagh, M.}, \emph {et~al.},\ }\bibfield  {title} {\bibinfo {title} {Randomized benchmarking of a high-fidelity remote cnot gate over a meter-scale microwave interconnect},\ }\href {https://arxiv.org/abs/2502.15034} {\bibfield  {journal} {\bibinfo  {journal} {arXiv:2502.15034}\ } (\bibinfo {year} {2025})}\BibitemShut {NoStop}%
\bibitem [{\citenamefont {Hashim}\ \emph {et~al.}(2024)Hashim, A. and Nguyen, L. B. and Goss, N. and Marinelli, B. and Naik, R. K. and Chistolini, T. and Hines, J. and Marceaux, J. P. and Kim, Y. and Gokhale, P. and Tomesh, T. and Chen, S. and Jiang, L. and Ferracin, S. and Rudinger, K. and Proctor, T. and Young, K. C. and Blume-Kohout, R. and Siddiqi, I.}]{Hashim2024a}%
  \BibitemOpen
  \bibfield  {author} {\bibinfo {author} {Hashim, A.}, \bibinfo {author} {Nguyen, L.~B.}, \bibinfo {author} {Goss, N.}, \emph {et~al.},\ }\bibfield  {title} {\bibinfo {title} {A practical introduction to benchmarking and characterization of quantum computers},\ }\href {https://arxiv.org/abs/2408.12064} {\bibfield  {journal} {\bibinfo  {journal} {arXiv:2408.12064}\ } (\bibinfo {year} {2024})}\BibitemShut {NoStop}%
\bibitem [{\citenamefont {Cross}\ \emph {et~al.}(2019)Cross, Andrew W. and Bishop, Lev S. and Sheldon, Sarah and Nation, Paul D. and Gambetta, Jay M.}]{Cross2018}%
  \BibitemOpen
  \bibfield  {author} {\bibinfo {author} {Cross, A.~W.}, \bibinfo {author} {Bishop, L.~S.}, \bibinfo {author} {Sheldon, S.}, \emph {et~al.},\ }\bibfield  {title} {\bibinfo {title} {Validating quantum computers using randomized model circuits},\ }\href {https://doi.org/10.1103/PhysRevA.100.032328} {\bibfield  {journal} {\bibinfo  {journal} {Phys. Rev. A}\ }\textbf {\bibinfo {volume} {100}},\ \bibinfo {pages} {032328} (\bibinfo {year} {2019})}\BibitemShut {NoStop}%
\bibitem [{\citenamefont {Proctor}\ \emph {et~al.}(2022)Proctor, Timothy and Seritan, Stefan and Rudinger, Kenneth and Nielsen, Erik and Blume-Kohout, Robin and Young, Kevin}]{Proctor2022}%
  \BibitemOpen
  \bibfield  {author} {\bibinfo {author} {Proctor, T.}, \bibinfo {author} {Seritan, S.}, \bibinfo {author} {Rudinger, K.}, \emph {et~al.},\ }\bibfield  {title} {\bibinfo {title} {Scalable randomized benchmarking of quantum computers using mirror circuits},\ }\href {https://doi.org/10.1103/PhysRevLett.129.150502} {\bibfield  {journal} {\bibinfo  {journal} {Phys. Rev. Lett.}\ }\textbf {\bibinfo {volume} {129}},\ \bibinfo {pages} {150502} (\bibinfo {year} {2022})}\BibitemShut {NoStop}%
\bibitem [{\citenamefont {Proctor}\ \emph {et~al.}(2025)Proctor, Timothy and Young, Kevin and Baczewski, Andrew D. and Blume-Kohout, Robin}]{Proctor2025}%
  \BibitemOpen
  \bibfield  {author} {\bibinfo {author} {Proctor, T.}, \bibinfo {author} {Young, K.}, \bibinfo {author} {Baczewski, A.~D.},\ and\ \bibinfo {author} {Blume-Kohout, R.},\ }\bibfield  {title} {\bibinfo {title} {Benchmarking quantum computers},\ }\href {https://doi.org/10.1038/s42254-024-00796-z} {\bibfield  {journal} {\bibinfo  {journal} {Nature Reviews Physics}\ }\textbf {\bibinfo {volume} {7}},\ \bibinfo {pages} {105} (\bibinfo {year} {2025})}\BibitemShut {NoStop}%
\bibitem [{\citenamefont {Elben}\ \emph {et~al.}(2020)Elben, Andreas and Vermersch, Beno\^{\i}t and van Bijnen, Rick and Kokail, Christian and Brydges, Tiff and Maier, Christine and Joshi, Manoj K. and Blatt, Rainer and Roos, Christian F. and Zoller, Peter}]{Elben2019}%
  \BibitemOpen
  \bibfield  {author} {\bibinfo {author} {Elben, A.}, \bibinfo {author} {Vermersch, B.}, \bibinfo {author} {van Bijnen, R.}, \emph {et~al.},\ }\bibfield  {title} {\bibinfo {title} {Cross-platform verification of intermediate scale quantum devices},\ }\href {https://doi.org/10.1103/PhysRevLett.124.010504} {\bibfield  {journal} {\bibinfo  {journal} {Phys. Rev. Lett.}\ }\textbf {\bibinfo {volume} {124}},\ \bibinfo {pages} {010504} (\bibinfo {year} {2020})}\BibitemShut {NoStop}%
\bibitem [{\citenamefont {Anshu}\ \emph {et~al.}(2022)Anshu, Anurag and Landau, Zeph and Liu, Yunchao}]{Anshu2022}%
  \BibitemOpen
  \bibfield  {author} {\bibinfo {author} {Anshu, A.}, \bibinfo {author} {Landau, Z.},\ and\ \bibinfo {author} {Liu, Y.},\ }\bibfield  {title} {\bibinfo {title} {Distributed quantum inner product estimation},\ }in\ \href {https://doi.org/10.1145/3519935.3519974} {\emph {\bibinfo {booktitle} {Proceedings of the 54th Annual ACM SIGACT Symposium on Theory of Computing}}},\ \bibinfo {series and number} {STOC ’22}\ (\bibinfo  {publisher} {ACM},\ \bibinfo {year} {2022})\ pp.\ \bibinfo {pages} {44--51}\BibitemShut {NoStop}%
\bibitem [{\citenamefont {Knörzer}\ \emph {et~al.}(2023)Knörzer, J. and Malz, D. and Cirac, J. I.}]{Knoerzer2023}%
  \BibitemOpen
  \bibfield  {author} {\bibinfo {author} {Knörzer, J.}, \bibinfo {author} {Malz, D.},\ and\ \bibinfo {author} {Cirac, J.~I.},\ }\bibfield  {title} {\bibinfo {title} {Cross-platform verification in quantum networks},\ }\href {https://doi.org/10.1103/physreva.107.062424} {\bibfield  {journal} {\bibinfo  {journal} {Physical Review A}\ }\textbf {\bibinfo {volume} {107}},\ \bibinfo {pages} {062424} (\bibinfo {year} {2023})}\BibitemShut {NoStop}%
\bibitem [{\citenamefont {Denzler}\ \emph {et~al.}(2025)Janek Denzler and Santiago Varona and Tommaso Guaita and Jose Carrasco}]{Denzler2025}%
  \BibitemOpen
  \bibfield  {author} {\bibinfo {author} {Denzler, J.}, \bibinfo {author} {Varona, S.}, \bibinfo {author} {Guaita, T.},\ and\ \bibinfo {author} {Carrasco, J.},\ }\bibfield  {title} {\bibinfo {title} {Highly-entangled, highly-doped states that are efficiently cross-device verifiable},\ }\href {https://arxiv.org/abs/2501.11688} {\bibfield  {journal} {\bibinfo  {journal} {arXiv:2501.11688}\ } (\bibinfo {year} {2025})}\BibitemShut {NoStop}%
\bibitem [{\citenamefont {Hinsche}\ \emph {et~al.}(2024)Hinsche, Marcel and Ioannou, Marios and Jerbi, Sofiene and Leone, Lorenzo and Eisert, Jens and Carrasco, Jose}]{Hinsche2024}%
  \BibitemOpen
  \bibfield  {author} {\bibinfo {author} {Hinsche, M.}, \bibinfo {author} {Ioannou, M.}, \bibinfo {author} {Jerbi, S.}, \emph {et~al.},\ }\bibfield  {title} {\bibinfo {title} {Efficient distributed inner product estimation via pauli sampling},\ }\href {https://doi.org/10.48550/ARXIV.2405.06544} {\bibfield  {journal} {\bibinfo  {journal} {arXiv:2405.06544}\ } (\bibinfo {year} {2024})}\BibitemShut {NoStop}%
\bibitem [{\citenamefont {Buhrman}\ \emph {et~al.}(2001)Buhrman, Harry and Cleve, Richard and Watrous, John and de Wolf, Ronald}]{Buhrman2001}%
  \BibitemOpen
  \bibfield  {author} {\bibinfo {author} {Buhrman, H.}, \bibinfo {author} {Cleve, R.}, \bibinfo {author} {Watrous, J.},\ and\ \bibinfo {author} {de~Wolf, R.},\ }\bibfield  {title} {\bibinfo {title} {Quantum fingerprinting},\ }\href {https://doi.org/10.1103/PhysRevLett.87.167902} {\bibfield  {journal} {\bibinfo  {journal} {Phys. Rev. Lett.}\ }\textbf {\bibinfo {volume} {87}},\ \bibinfo {pages} {167902} (\bibinfo {year} {2001})}\BibitemShut {NoStop}%
\bibitem [{\citenamefont {Ekert}\ \emph {et~al.}(2002)Ekert, Artur K. and Alves, Carolina Moura and Oi, Daniel K. L. and Horodecki, Michał and Horodecki, Paweł and Kwek, L. C.}]{Ekert2002}%
  \BibitemOpen
  \bibfield  {author} {\bibinfo {author} {Ekert, A.~K.}, \bibinfo {author} {Alves, C.~M.}, \bibinfo {author} {Oi, D. K.~L.}, \emph {et~al.},\ }\bibfield  {title} {\bibinfo {title} {Direct estimations of linear and nonlinear functionals of a quantum state},\ }\href {https://doi.org/10.1103/physrevlett.88.217901} {\bibfield  {journal} {\bibinfo  {journal} {Physical Review Letters}\ }\textbf {\bibinfo {volume} {88}},\ \bibinfo {pages} {217901} (\bibinfo {year} {2002})}\BibitemShut {NoStop}%
\bibitem [{\citenamefont {Garcia-Escartin}\ and\ \citenamefont {Chamorro-Posada}(2013)Garcia-Escartin, Juan Carlos and Chamorro-Posada, Pedro}]{GarciaEscartin2013}%
  \BibitemOpen
  \bibfield  {author} {\bibinfo {author} {Garcia-Escartin, J.~C.}\ and\ \bibinfo {author} {Chamorro-Posada, P.},\ }\bibfield  {title} {\bibinfo {title} {swap test and hong-ou-mandel effect are equivalent},\ }\href {https://doi.org/10.1103/physreva.87.052330} {\bibfield  {journal} {\bibinfo  {journal} {Physical Review A}\ }\textbf {\bibinfo {volume} {87}},\ \bibinfo {pages} {052330} (\bibinfo {year} {2013})}\BibitemShut {NoStop}%
\bibitem [{\citenamefont {Cincio}\ \emph {et~al.}(2018)Cincio, Lukasz and Subaşı, Yiğit and Sornborger, Andrew T and Coles, Patrick J}]{Cincio2018}%
  \BibitemOpen
  \bibfield  {author} {\bibinfo {author} {Cincio, L.}, \bibinfo {author} {Subaşı, Y.}, \bibinfo {author} {Sornborger, A.~T.},\ and\ \bibinfo {author} {Coles, P.~J.},\ }\bibfield  {title} {\bibinfo {title} {Learning the quantum algorithm for state overlap},\ }\href {https://doi.org/10.1088/1367-2630/aae94a} {\bibfield  {journal} {\bibinfo  {journal} {New Journal of Physics}\ }\textbf {\bibinfo {volume} {20}},\ \bibinfo {pages} {113022} (\bibinfo {year} {2018})}\BibitemShut {NoStop}%
\bibitem [{\citenamefont {Zhu}\ \emph {et~al.}(2022)Zhu, D. and Cian, Z. and Noel, C. and Risinger, A. and Biswas, D. and Egan, L. and Zhu, Y. and Green, A. M. and Maksymov, A. and Nam, Y. and Cetina, M. and Linke, N. M. and Hafezi, M. and Monroe, C.}]{Zhu2022e}%
  \BibitemOpen
  \bibfield  {author} {\bibinfo {author} {Zhu, D.}, \bibinfo {author} {Cian, Z.}, \bibinfo {author} {Noel, C.}, \emph {et~al.},\ }\bibfield  {title} {\bibinfo {title} {Cross-platform comparison of arbitrary quantum computations},\ }\href {https://doi.org/https://doi.org/10.1038/s41467-022-34279-5} {\bibfield  {journal} {\bibinfo  {journal} {Nat. Commun.}\ }\textbf {\bibinfo {volume} {13}},\ \bibinfo {pages} {1} (\bibinfo {year} {2022})}\BibitemShut {NoStop}%
\bibitem [{\citenamefont {Zheng}\ \emph {et~al.}(2024)Zheng, Congcong and Yu, Xutao and Wang, Kun}]{Zheng2024a}%
  \BibitemOpen
  \bibfield  {author} {\bibinfo {author} {Zheng, C.}, \bibinfo {author} {Yu, X.},\ and\ \bibinfo {author} {Wang, K.},\ }\bibfield  {title} {\bibinfo {title} {Cross-platform comparison of arbitrary quantum processes},\ }\href {https://doi.org/10.1038/s41534-023-00797-3} {\bibfield  {journal} {\bibinfo  {journal} {npj Quantum Information}\ }\textbf {\bibinfo {volume} {10}},\ \bibinfo {pages} {4} (\bibinfo {year} {2024})}\BibitemShut {NoStop}%
\bibitem [{\citenamefont {Daley}\ \emph {et~al.}(2012)Daley, A. J. and Pichler, H. and Schachenmayer, J. and Zoller, P.}]{Daley2012}%
  \BibitemOpen
  \bibfield  {author} {\bibinfo {author} {Daley, A.~J.}, \bibinfo {author} {Pichler, H.}, \bibinfo {author} {Schachenmayer, J.},\ and\ \bibinfo {author} {Zoller, P.},\ }\bibfield  {title} {\bibinfo {title} {Measuring entanglement growth in quench dynamics of bosons in an optical lattice},\ }\href {https://doi.org/10.1103/physrevlett.109.020505} {\bibfield  {journal} {\bibinfo  {journal} {Physical Review Letters}\ }\textbf {\bibinfo {volume} {109}},\ \bibinfo {pages} {020505} (\bibinfo {year} {2012})}\BibitemShut {NoStop}%
\bibitem [{\citenamefont {Islam}\ \emph {et~al.}(2015)Islam, Rajibul and Ma, Ruichao and Preiss, Philipp M. and Eric Tai, M. and Lukin, Alexander and Rispoli, Matthew and Greiner, Markus}]{Islam2015}%
  \BibitemOpen
  \bibfield  {author} {\bibinfo {author} {Islam, R.}, \bibinfo {author} {Ma, R.}, \bibinfo {author} {Preiss, P.~M.}, \emph {et~al.},\ }\bibfield  {title} {\bibinfo {title} {Measuring entanglement entropy in a quantum many-body system},\ }\href {https://doi.org/10.1038/nature15750} {\bibfield  {journal} {\bibinfo  {journal} {Nature}\ }\textbf {\bibinfo {volume} {528}},\ \bibinfo {pages} {77} (\bibinfo {year} {2015})}\BibitemShut {NoStop}%
\bibitem [{\citenamefont {Bluvstein}\ \emph {et~al.}(2022)Dolev Bluvstein and Harry Levine and Giulia Semeghini and Tout T. Wang and Sepehr Ebadi and Marcin Kalinowski and Alexander Keesling and Nishad Maskara and Hannes Pichler and Markus Greiner and Vladan Vuleti{\'{c}} and Mikhail D. Lukin}]{Bluvstein2022}%
  \BibitemOpen
  \bibfield  {author} {\bibinfo {author} {Bluvstein, D.}, \bibinfo {author} {Levine, H.}, \bibinfo {author} {Semeghini, G.}, \emph {et~al.},\ }\bibfield  {title} {\bibinfo {title} {A quantum processor based on coherent transport of entangled atom arrays},\ }\href {https://doi.org/10.1038/s41586-022-04592-6} {\bibfield  {journal} {\bibinfo  {journal} {Nature}\ }\textbf {\bibinfo {volume} {604}},\ \bibinfo {pages} {451} (\bibinfo {year} {2022})}\BibitemShut {NoStop}%
\bibitem [{\citenamefont {Karamlou}\ \emph {et~al.}(2024)Karamlou, Amir H. and Rosen, Ilan T. and Muschinske, Sarah E. and Barrett, Cora N. and Di Paolo, Agustin and Ding, Leon and Harrington, Patrick M. and Hays, Max and Das, Rabindra and Kim, David K. and Niedzielski, Bethany M. and Schuldt, Meghan and Serniak, Kyle and Schwartz, Mollie E. and Yoder, Jonilyn L. and Gustavsson, Simon and Yanay, Yariv and Grover, Jeffrey A. and Oliver, William D.}]{Karamlou2024}%
  \BibitemOpen
  \bibfield  {author} {\bibinfo {author} {Karamlou, A.~H.}, \bibinfo {author} {Rosen, I.~T.}, \bibinfo {author} {Muschinske, S.~E.}, \emph {et~al.},\ }\bibfield  {title} {\bibinfo {title} {Probing entanglement in a 2d hard-core bose–hubbard lattice},\ }\href {https://doi.org/10.1038/s41586-024-07325-z} {\bibfield  {journal} {\bibinfo  {journal} {Nature}\ }\textbf {\bibinfo {volume} {629}},\ \bibinfo {pages} {561} (\bibinfo {year} {2024})}\BibitemShut {NoStop}%
\bibitem [{\citenamefont {Kosen}\ \emph {et~al.}(2024)Kosen, S. and Li, H. and Rommel, M. and Rehammar, R. and Caputo, M. and Gronberg, L. and Fernandez-Pendas, J. and Kockum, A. F. and Biznarova, J. and Chen, L. and Krizan, C. and Nylander, A. and Osman, A. and Roudsari, A. F. and Shiri, D. and Tancredi, G. and Govenius, J. and Bylander, J.}]{Kosen2024}%
  \BibitemOpen
  \bibfield  {author} {\bibinfo {author} {Kosen, S.}, \bibinfo {author} {Li, H.}, \bibinfo {author} {Rommel, M.}, \emph {et~al.},\ }\bibfield  {title} {\bibinfo {title} {Signal crosstalk in a flip-chip quantum processor},\ }\href {https://arxiv.org/abs/2403.00285} {\bibfield  {journal} {\bibinfo  {journal} {arXiv:2403.00285}\ } (\bibinfo {year} {2024})}\BibitemShut {NoStop}%
\bibitem [{\citenamefont {Smith}\ \emph {et~al.}(2023)Smith, Kaitlin N. and Ravi, Gokul Subramanian and Baker, Jonathan M. and Chong, Frederic T.}]{Smith2022a}%
  \BibitemOpen
  \bibfield  {author} {\bibinfo {author} {Smith, K.~N.}, \bibinfo {author} {Ravi, G.~S.}, \bibinfo {author} {Baker, J.~M.},\ and\ \bibinfo {author} {Chong, F.~T.},\ }\bibfield  {title} {\bibinfo {title} {Scaling superconducting quantum computers with chiplet architectures},\ }in\ \href {https://doi.org/10.1109/MICRO56248.2022.00078} {\emph {\bibinfo {booktitle} {Proceedings of the 55th Annual IEEE/ACM International Symposium on Microarchitecture}}},\ \bibinfo {series and number} {MICRO '22}\ (\bibinfo  {publisher} {IEEE Press},\ \bibinfo {year} {2023})\ p.\ \bibinfo {pages} {1092–1109}\BibitemShut {NoStop}%
\bibitem [{\citenamefont {Kim}\ \emph {et~al.}(2023)Kim, Youngseok and Eddins, Andrew and Anand, Sajant and Wei, Ken Xuan and van den Berg, Ewout and Rosenblatt, Sami and Nayfeh, Hasan and Wu, Yantao and Zaletel, Michael and Temme, Kristan and Kandala, Abhinav}]{Kim2023b}%
  \BibitemOpen
  \bibfield  {author} {\bibinfo {author} {Kim, Y.}, \bibinfo {author} {Eddins, A.}, \bibinfo {author} {Anand, S.}, \emph {et~al.},\ }\bibfield  {title} {\bibinfo {title} {Evidence for the utility of quantum computing before fault tolerance},\ }\href {https://doi.org/10.1038/s41586-023-06096-3} {\bibfield  {journal} {\bibinfo  {journal} {Nature}\ }\textbf {\bibinfo {volume} {618}},\ \bibinfo {pages} {500} (\bibinfo {year} {2023})}\BibitemShut {NoStop}%
\bibitem [{\citenamefont {Acharya}\ \emph {et~al.}(2025{\natexlab{a}})Acharya, Rajeev and Abanin, Dmitry A. and Aghababaie-Beni, Laleh and Aleiner, Igor and Andersen, Trond I. and Ansmann, Markus and Arute, Frank and Arya, Kunal and Asfaw, Abraham and Astrakhantsev, Nikita and Atalaya, Juan and Babbush, Ryan and Bacon, Dave and Ballard, Brian and Bardin, Joseph C. and Bausch, Johannes and Bengtsson, Andreas and Bilmes, Alexander and Blackwell, Sam and Boixo, Sergio and Bortoli, Gina and Bourassa, Alexandre and Bovaird, Jenna and Brill, Leon and Broughton, Michael and Browne, David A. and Buchea, Brett and Buckley, Bob B. and Buell, David A. and Burger, Tim and Burkett, Brian and Bushnell, Nicholas and Cabrera, Anthony and Campero, Juan and Chang, Hung-Shen and Chen, Yu and Chen, Zijun and Chiaro, Ben and Chik, Desmond and Chou, Charina and Claes, Jahan and Cleland, Agnetta Y. and Cogan, Josh and Collins, Roberto and Conner, Paul and Courtney, William and Crook, Alexander L. and Curtin, Ben and Das, Sayan and
  Davies, Alex and De Lorenzo, Laura and Debroy, Dripto M. and Demura, Sean and Devoret, Michel and Di Paolo, Agustin and Donohoe, Paul and Drozdov, Ilya and Dunsworth, Andrew and Earle, Clint and Edlich, Thomas and Eickbusch, Alec and Elbag, Aviv Moshe and Elzouka, Mahmoud and Erickson, Catherine and Faoro, Lara and Farhi, Edward and Ferreira, Vinicius S. and Burgos, Leslie Flores and Forati, Ebrahim and Fowler, Austin G. and Foxen, Brooks and Ganjam, Suhas and Garcia, Gonzalo and Gasca, Robert and Genois, {\'E}lie and Giang, William and Gidney, Craig and Gilboa, Dar and Gosula, Raja and Dau, Alejandro Grajales and Graumann, Dietrich and Greene, Alex and Gross, Jonathan A. and Habegger, Steve and Hall, John and Hamilton, Michael C. and Hansen, Monica and Harrigan, Matthew P. and Harrington, Sean D. and Heras, Francisco J. H. and Heslin, Stephen and Heu, Paula and Higgott, Oscar and Hill, Gordon and Hilton, Jeremy and Holland, George and Hong, Sabrina and Huang, Hsin-Yuan and Huff, Ashley and Huggins, William
  J. and Ioffe, Lev B. and Isakov, Sergei V. and Iveland, Justin and Jeffrey, Evan and Jiang, Zhang and Jones, Cody and Jordan, Stephen and Joshi, Chaitali and Juhas, Pavol and Kafri, Dvir and Kang, Hui and Karamlou, Amir H. and Kechedzhi, Kostyantyn and Kelly, Julian and Khaire, Trupti and Khattar, Tanuj and Khezri, Mostafa and Kim, Seon and Klimov, Paul V. and Klots, Andrey R. and Kobrin, Bryce and Kohli, Pushmeet and Korotkov, Alexander N. and Kostritsa, Fedor and Kothari, Robin and Kozlovskii, Borislav and Kreikebaum, John Mark and Kurilovich, Vladislav D. and Lacroix, Nathan and Landhuis, David and Lange-Dei, Tiano and Langley, Brandon W. and Laptev, Pavel and Lau, Kim-Ming and Le Guevel, Lo{\"i}ck and Ledford, Justin and Lee, Joonho and Lee, Kenny and Lensky, Yuri D. and Leon, Shannon and Lester, Brian J. and Li, Wing Yan and Li, Yin and Lill, Alexander T. and Liu, Wayne and Livingston, William P. and Locharla, Aditya and Lucero, Erik and Lundahl, Daniel and Lunt, Aaron and Madhuk, Sid and Malone, Fionn
  D. and Maloney, Ashley and Mandr{\`a}, Salvatore and Manyika, James and Martin, Leigh S. and Martin, Orion and Martin, Steven and Maxfield, Cameron and McClean, Jarrod R. and McEwen, Matt and Meeks, Seneca and Megrant, Anthony and Mi, Xiao and Miao, Kevin C. and Mieszala, Amanda and Molavi, Reza and Molina, Sebastian and Montazeri, Shirin and Morvan, Alexis and Movassagh, Ramis and Mruczkiewicz, Wojciech and Naaman, Ofer and Neeley, Matthew and Neill, Charles and Nersisyan, Ani and Neven, Hartmut and Newman, Michael and Ng, Jiun How and Nguyen, Anthony and Nguyen, Murray and Ni, Chia-Hung and Niu, Murphy Yuezhen and O'Brien, Thomas E. and Oliver, William D. and Opremcak, Alex and Ottosson, Kristoffer and Petukhov, Andre and Pizzuto, Alex and Platt, John and Potter, Rebecca and Pritchard, Orion and Pryadko, Leonid P. and Quintana, Chris and Ramachandran, Ganesh and Reagor, Matthew J. and Redding, John and Rhodes, David M. and Roberts, Gabrielle and Rosenberg, Eliott and Rosenfeld, Emma and Roushan, Pedram
  and Rubin, Nicholas C. and Saei, Negar and Sank, Daniel and Sankaragomathi, Kannan and Satzinger, Kevin J. and Schurkus, Henry F. and Schuster, Christopher and Senior, Andrew W. and Shearn, Michael J. and Shorter, Aaron and Shutty, Noah and Shvarts, Vladimir and Singh, Shraddha and Sivak, Volodymyr and Skruzny, Jindra and Small, Spencer and Smelyanskiy, Vadim and Smith, W. Clarke and Somma, Rolando D. and Springer, Sofia and Sterling, George and Strain, Doug and Suchard, Jordan and Szasz, Aaron and Sztein, Alex and Thor, Douglas and Torres, Alfredo and Torunbalci, M. Mert and Vaishnav, Abeer and Vargas, Justin and Vdovichev, Sergey and Vidal, Guifre and Villalonga, Benjamin and Heidweiller, Catherine Vollgraff and Waltman, Steven and Wang, Shannon X. and Ware, Brayden and Weber, Kate and Weidel, Travis and White, Theodore and Wong, Kristi and Woo, Bryan W. K. and Xing, Cheng and Yao, Z. Jamie and Yeh, Ping and Ying, Bicheng and Yoo, Juhwan and Yosri, Noureldin and Young, Grayson and Zalcman, Adam and Zhang,
  Yaxing and Zhu, Ningfeng and Zobrist, Nicholas and AI, Google Quantum and {Collaborators}}]{Acharya2025}%
  \BibitemOpen
  \bibfield  {author} {\bibinfo {author} {Acharya, R.}, \bibinfo {author} {Abanin, D.~A.}, \bibinfo {author} {Aghababaie-Beni, L.}, \emph {et~al.},\ }\bibfield  {title} {\bibinfo {title} {Quantum error correction below the surface code threshold},\ }\href {https://doi.org/10.1038/s41586-024-08449-y} {\bibfield  {journal} {\bibinfo  {journal} {Nature}\ }\textbf {\bibinfo {volume} {638}},\ \bibinfo {pages} {920} (\bibinfo {year} {2025}{\natexlab{a}})}\BibitemShut {NoStop}%
\bibitem [{\citenamefont {Gao}\ \emph {et~al.}(2025)Gao, Dongxin and Fan, Daojin and Zha, Chen and Bei, Jiahao and Cai, Guoqing and Cai, Jianbin and Cao, Sirui and Chen, Fusheng and Chen, Jiang and Chen, Kefu and Chen, Xiawei and Chen, Xiqing and Chen, Zhe and Chen, Zhiyuan and Chen, Zihua and Chu, Wenhao and Deng, Hui and Deng, Zhibin and Ding, Pei and Ding, Xun and Ding, Zhuzhengqi and Dong, Shuai and Dong, Yupeng and Fan, Bo and Fu, Yuanhao and Gao, Song and Ge, Lei and Gong, Ming and Gui, Jiacheng and Guo, Cheng and Guo, Shaojun and Guo, Xiaoyang and Han, Lianchen and He, Tan and Hong, Linyin and Hu, Yisen and Huang, He-Liang and Huo, Yong-Heng and Jiang, Tao and Jiang, Zuokai and Jin, Honghong and Leng, Yunxiang and Li, Dayu and Li, Dongdong and Li, Fangyu and Li, Jiaqi and Li, Jinjin and Li, Junyan and Li, Junyun and Li, Na and Li, Shaowei and Li, Wei and Li, Yuhuai and Li, Yuan and Liang, Futian and Liang, Xuelian and Liao, Nanxing and Lin, Jin and Lin, Weiping and Liu, Dailin and Liu, Hongxiu and Liu,
  Maliang and Liu, Xinyu and Liu, Xuemeng and Liu, Yancheng and Lou, Haoxin and Ma, Yuwei and Meng, Lingxin and Mou, Hao and Nan, Kailiang and Nie, Binghan and Nie, Meijuan and Ning, Jie and Niu, Le and Peng, Wenyi and Qian, Haoran and Rong, Hao and Rong, Tao and Shen, Huiyan and Shen, Qiong and Su, Hong and Su, Feifan and Sun, Chenyin and Sun, Liangchao and Sun, Tianzuo and Sun, Yingxiu and Tan, Yimeng and Tan, Jun and Tang, Longyue and Tu, Wenbing and Wan, Cai and Wang, Jiafei and Wang, Biao and Wang, Chang and Wang, Chen and Wang, Chu and Wang, Jian and Wang, Liangyuan and Wang, Rui and Wang, Shengtao and Wang, Xiaomin and Wang, Xinzhe and Wang, Xunxun and Wang, Yeru and Wei, Zuolin and Wei, Jiazhou and Wu, Dachao and Wu, Gang and Wu, Jin and Wu, Shengjie and Wu, Yulin and Xie, Shiyong and Xin, Lianjie and Xu, Yu and Xue, Chun and Yan, Kai and Yang, Weifeng and Yang, Xinpeng and Yang, Yang and Ye, Yangsen and Ye, Zhenping and Ying, Chong and Yu, Jiale and Yu, Qinjing and Yu, Wenhu and Zeng, Xiangdong and
  Zhan, Shaoyu and Zhang, Feifei and Zhang, Haibin and Zhang, Kaili and Zhang, Pan and Zhang, Wen and Zhang, Yiming and Zhang, Yongzhuo and Zhang, Lixiang and Zhao, Guming and Zhao, Peng and Zhao, Xianhe and Zhao, Xintao and Zhao, Youwei and Zhao, Zhong and Zheng, Luyuan and Zhou, Fei and Zhou, Liang and Zhou, Na and Zhou, Naibin and Zhou, Shifeng and Zhou, Shuang and Zhou, Zhengxiao and Zhu, Chengjun and Zhu, Qingling and Zou, Guihong and Zou, Haonan and Zhang, Qiang and Lu, Chao-Yang and Peng, Cheng-Zhi and Zhu, Xiaobo and Pan, Jian-Wei}]{Gao2025}%
  \BibitemOpen
  \bibfield  {author} {\bibinfo {author} {Gao, D.}, \bibinfo {author} {Fan, D.}, \bibinfo {author} {Zha, C.}, \emph {et~al.},\ }\bibfield  {title} {\bibinfo {title} {Establishing a new benchmark in quantum computational advantage with 105-qubit zuchongzhi 3.0 processor},\ }\href {http://dx.doi.org/10.1103/PhysRevLett.134.090601} {\bibfield  {journal} {\bibinfo  {journal} {Phys. Rev. Lett.}\ }\textbf {\bibinfo {volume} {134}} (\bibinfo {year} {2025})}\BibitemShut {NoStop}%
\bibitem [{\citenamefont {Bravyi}\ \emph {et~al.}(2022)Bravyi, Sergey and Dial, Oliver and Gambetta, Jay M. and Gil, Darío and Nazario, Zaira}]{Bravyi2022a}%
  \BibitemOpen
  \bibfield  {author} {\bibinfo {author} {Bravyi, S.}, \bibinfo {author} {Dial, O.}, \bibinfo {author} {Gambetta, J.~M.}, \emph {et~al.},\ }\bibfield  {title} {\bibinfo {title} {{The future of quantum computing with superconducting qubits}},\ }\href {https://doi.org/10.1063/5.0082975} {\bibfield  {journal} {\bibinfo  {journal} {Journal of Applied Physics}\ }\textbf {\bibinfo {volume} {132}},\ \bibinfo {pages} {160902} (\bibinfo {year} {2022})}\BibitemShut {NoStop}%
\bibitem [{\citenamefont {Norris}\ \emph {et~al.}(2025)Graham J. Norris and Kieran Dalton and Dante Colao Zanuz and Alexander Rommens and Alexander Flasby and Mohsen Bahrami Panah and François Swiadek and Colin Scarato and Christoph Hellings and Jean-Claude Besse and Andreas Wallraff}]{Norris2025}%
  \BibitemOpen
  \bibfield  {author} {\bibinfo {author} {Norris, G.~J.}, \bibinfo {author} {Dalton, K.}, \bibinfo {author} {Zanuz, D.~C.}, \emph {et~al.},\ }\bibfield  {title} {\bibinfo {title} {Performance characterization of a multi-module quantum processor with static inter-chip couplers},\ }\href {https://arxiv.org/abs/2503.12603} {\bibfield  {journal} {\bibinfo  {journal} {arXiv:2503.12603}\ } (\bibinfo {year} {2025})}\BibitemShut {NoStop}%
\bibitem [{\citenamefont {Ihssen}\ \emph {et~al.}(2025)Ihssen, Soeren and Geisert, Simon and Jauma, Gabriel and Winkel, Patrick and Spiecker, Martin and Zapata, Nicolas and Gosling, Nicolas and Paluch, Patrick and Pino, Manuel and Reisinger, Thomas and Wernsdorfer, Wolfgang and Garcia-Ripoll, Juan Jose and Pop, Ioan M.}]{Ihssen2025}%
  \BibitemOpen
  \bibfield  {author} {\bibinfo {author} {Ihssen, S.}, \bibinfo {author} {Geisert, S.}, \bibinfo {author} {Jauma, G.}, \emph {et~al.},\ }\bibfield  {title} {\bibinfo {title} {Low crosstalk modular flip-chip architecture for coupled superconducting qubits},\ }\href {https://doi.org/10.1063/5.0245667} {\bibfield  {journal} {\bibinfo  {journal} {Applied Physics Letters}\ }\textbf {\bibinfo {volume} {126}},\ \bibinfo {pages} {134003} (\bibinfo {year} {2025})}\BibitemShut {NoStop}%
\bibitem [{\citenamefont {Strauch}\ \emph {et~al.}(2003)Strauch, Frederick W. and Johnson, Philip R. and Dragt, Alex J. and Lobb, C. J. and Anderson, J. R. and Wellstood, F. C.}]{Strauch2003}%
  \BibitemOpen
  \bibfield  {author} {\bibinfo {author} {Strauch, F.~W.}, \bibinfo {author} {Johnson, P.~R.}, \bibinfo {author} {Dragt, A.~J.}, \emph {et~al.},\ }\bibfield  {title} {\bibinfo {title} {Quantum logic gates for coupled superconducting phase qubits},\ }\href {https://doi.org/10.1103/PhysRevLett.91.167005} {\bibfield  {journal} {\bibinfo  {journal} {Phys. Rev. Lett.}\ }\textbf {\bibinfo {volume} {91}},\ \bibinfo {pages} {167005} (\bibinfo {year} {2003})}\BibitemShut {NoStop}%
\bibitem [{\citenamefont {DiCarlo}\ \emph {et~al.}(2009)DiCarlo, L. and Chow, J. M. and Gambetta, J. M. and Bishop, Lev S. and Johnson, B. R. and Schuster, D. I. and Majer, J. and Blais, A. and Frunzio, L. and Girvin, S. M. and Schoelkopf, R. J.}]{DiCarlo2009}%
  \BibitemOpen
  \bibfield  {author} {\bibinfo {author} {DiCarlo, L.}, \bibinfo {author} {Chow, J.~M.}, \bibinfo {author} {Gambetta, J.~M.}, \emph {et~al.},\ }\bibfield  {title} {\bibinfo {title} {Demonstration of two-qubit algorithms with a superconducting quantum processor},\ }\href {https://doi.org/10.1038/nature08121} {\bibfield  {journal} {\bibinfo  {journal} {Nature}\ }\textbf {\bibinfo {volume} {460}},\ \bibinfo {pages} {240} (\bibinfo {year} {2009})}\BibitemShut {NoStop}%
\bibitem [{\citenamefont {Hellings}\ \emph {et~al.}(2025)Hellings, Christoph and Lacroix, Nathan and Remm, Ants and Boell, Richard and Herrmann, Johannes and Laz\u{a}r, Stefania and Krinner, Sebastian and Swiadek, Fran\c{c}ois and Andersen, Christian Kraglund and Eichler, Christopher and Wallraff, Andreas}]{Hellings2025}%
  \BibitemOpen
  \bibfield  {author} {\bibinfo {author} {Hellings, C.}, \bibinfo {author} {Lacroix, N.}, \bibinfo {author} {Remm, A.}, \emph {et~al.},\ }\bibfield  {title} {\bibinfo {title} {Calibrating magnetic flux control in superconducting circuits by compensating distortions on time scales from nanoseconds up to tens of microseconds},\ }\href {https://arxiv.org/abs/2503.04610} {\bibfield  {journal} {\bibinfo  {journal} {arXiv:2503.04610}\ } (\bibinfo {year} {2025})}\BibitemShut {NoStop}%
\bibitem [{\citenamefont {Walter}\ \emph {et~al.}(2017)Walter, T. and Kurpiers, P. and Gasparinetti, S. and Magnard, P. and Poto\v{c}nik, A. and Salath\'e, Y. and Pechal, M. and Mondal, M. and Oppliger, M. and Eichler, C. and Wallraff, A.}]{Walter2017}%
  \BibitemOpen
  \bibfield  {author} {\bibinfo {author} {Walter, T.}, \bibinfo {author} {Kurpiers, P.}, \bibinfo {author} {Gasparinetti, S.}, \emph {et~al.},\ }\bibfield  {title} {\bibinfo {title} {Rapid high-fidelity single-shot dispersive readout of superconducting qubits},\ }\href {https://doi.org/10.1103/PhysRevApplied.7.054020} {\bibfield  {journal} {\bibinfo  {journal} {Phys. Rev. Appl.}\ }\textbf {\bibinfo {volume} {7}},\ \bibinfo {pages} {054020} (\bibinfo {year} {2017})}\BibitemShut {NoStop}%
\bibitem [{\citenamefont {Heinsoo}\ \emph {et~al.}(2018)Heinsoo, Johannes and Andersen, Christian Kraglund and Remm, Ants and Krinner, Sebastian and Walter, Theodore and Salath\'{e}, Yves and Gasparinetti, Simone and Besse, Jean-Claude and Poto\v{c}nik, Anton and Wallraff, Andreas and Eichler, Christopher}]{Heinsoo2018}%
  \BibitemOpen
  \bibfield  {author} {\bibinfo {author} {Heinsoo, J.}, \bibinfo {author} {Andersen, C.~K.}, \bibinfo {author} {Remm, A.}, \emph {et~al.},\ }\bibfield  {title} {\bibinfo {title} {Rapid high-fidelity multiplexed readout of superconducting qubits},\ }\href {https://doi.org/10.1103/PhysRevApplied.10.034040} {\bibfield  {journal} {\bibinfo  {journal} {Phys. Rev. Appl.}\ }\textbf {\bibinfo {volume} {10}},\ \bibinfo {pages} {034040} (\bibinfo {year} {2018})}\BibitemShut {NoStop}%
\bibitem [{\citenamefont {Norris}\ \emph {et~al.}(2024)Norris, Graham J. and Michaud, Laurent and Pahl, David and Kerschbaum, Michael and Eichler, Christopher and Besse, Jean-Claude and Wallraff, Andreas}]{Norris2024}%
  \BibitemOpen
  \bibfield  {author} {\bibinfo {author} {Norris, G.~J.}, \bibinfo {author} {Michaud, L.}, \bibinfo {author} {Pahl, D.}, \emph {et~al.},\ }\bibfield  {title} {\bibinfo {title} {Improved parameter targeting in {3D}-integrated superconducting circuits through a polymer spacer process},\ }\href {https://doi.org/10.1140/epjqt/s40507-023-00213-x} {\bibfield  {journal} {\bibinfo  {journal} {EPJ Quantum Technol.}\ }\textbf {\bibinfo {volume} {11}},\ \bibinfo {pages} {5} (\bibinfo {year} {2024})}\BibitemShut {NoStop}%
\bibitem [{\citenamefont {Krinner}\ \emph {et~al.}(2022)Sebastian Krinner and Nathan Lacroix and Ants Remm and Agustin Di Paolo and Elie Genois and Catherine Leroux and Christoph Hellings and Stefania Laz\u{a}r and Francois Swiadek and Johannes Herrmann and Graham J. Norris and Christian Kraglund Andersen and Markus M\"{u}ller and Alexandre Blais and Christopher Eichler and Andreas Wallraff}]{Krinner2022}%
  \BibitemOpen
  \bibfield  {author} {\bibinfo {author} {Krinner, S.}, \bibinfo {author} {Lacroix, N.}, \bibinfo {author} {Remm, A.}, \emph {et~al.},\ }\bibfield  {title} {\bibinfo {title} {Realizing repeated quantum error correction in a distance-three surface code},\ }\href {https://doi.org/10.1038/s41586-022-04566-8} {\bibfield  {journal} {\bibinfo  {journal} {Nature}\ }\textbf {\bibinfo {volume} {605}},\ \bibinfo {pages} {669} (\bibinfo {year} {2022})}\BibitemShut {NoStop}%
\bibitem [{\citenamefont {Besedin}\ \emph {et~al.}(2025)Besedin, Ilya and Kerschbaum, Michael and Knoll, Jonathan and Hesner, Ian and B\"{o}deker, Lukas and Colmenarez, Luis and Hofele, Luca and Lacroix, Nathan and Hellings, Christoph and Swiadek, Fran\c{c}ois and Flasby, Alexander and Panah, Mohsen Bahrami and Zanuz, Dante Colao and M\"{u}ller, Markus and Wallraff, Andreas}]{Besedin2025}%
  \BibitemOpen
  \bibfield  {author} {\bibinfo {author} {Besedin, I.}, \bibinfo {author} {Kerschbaum, M.}, \bibinfo {author} {Knoll, J.}, \emph {et~al.},\ }\bibfield  {title} {\bibinfo {title} {Realizing lattice surgery on two distance-three repetition codes with superconducting qubits},\ }\href {https://arxiv.org/abs/2501.04612} {\bibfield  {journal} {\bibinfo  {journal} {arXiv:2501.04612}\ } (\bibinfo {year} {2025})}\BibitemShut {NoStop}%
\bibitem [{\citenamefont {Kosen}\ \emph {et~al.}(2022)Sandoko Kosen and Hang-Xi Li and Marcus Rommel and Daryoush Shiri and Christopher Warren and Leif Gr\"{o}nberg and Jaakko Salonen and Tahereh Abad and Janka Bizn\'{a}rov\'{a} and Marco Caputo and Liangyu Chen and Kestutis Grigoras and G\"{o}ran Johansson and Anton Frisk Kockum and Christian Kri\v{z}an and Daniel P\'{e}rez Lozano and Graham J. Norris and Amr Osman and Jorge Fern\'{a}ndez-Pend\'{a}s and Alberto Ronzani and Anita Fadavi Roudsari and Slawomir Simbierowicz and Giovanna Tancredi and Andreas Wallraff and Christopher Eichler and Joonas Govenius and Jonas Bylander}]{Kosen2022}%
  \BibitemOpen
  \bibfield  {author} {\bibinfo {author} {Kosen, S.}, \bibinfo {author} {Li, H.-X.}, \bibinfo {author} {Rommel, M.}, \emph {et~al.},\ }\bibfield  {title} {\bibinfo {title} {Building blocks of a flip-chip integrated superconducting quantum processor},\ }\href {https://doi.org/10.1088/2058-9565/ac734b} {\bibfield  {journal} {\bibinfo  {journal} {Quantum Sci. Technol.}\ }\textbf {\bibinfo {volume} {7}},\ \bibinfo {pages} {035018} (\bibinfo {year} {2022})}\BibitemShut {NoStop}%
\bibitem [{\citenamefont {Scarato}\ \emph {et~al.}(2025)Colin Scarato and Kilian Hanke and Ants Remm and Stefania Laz\u{a}r and Nathan Lacroix and Dante Colao Zanuz and Alexander Flasby and Andreas Wallraff and Christoph Hellings}]{Scarato2025}%
  \BibitemOpen
  \bibfield  {author} {\bibinfo {author} {Scarato, C.}, \bibinfo {author} {Hanke, K.}, \bibinfo {author} {Remm, A.}, \emph {et~al.},\ }\bibfield  {title} {\bibinfo {title} {{Realizing a Continuous Set of Two-Qubit Gates Parameterized by an Idle Time}},\ }\href {https://arxiv.org/abs/2503.11204} {\bibfield  {journal} {\bibinfo  {journal} {arXiv:2503.11204}\ } (\bibinfo {year} {2025})}\BibitemShut {NoStop}%
\bibitem [{\citenamefont {Miao}\ \emph {et~al.}(2023)Miao, Kevin C. and McEwen, Matt and Atalaya, Juan and Kafri, Dvir and Pryadko, Leonid P. and Bengtsson, Andreas and Opremcak, Alex and Satzinger, Kevin J. and Chen, Zijun and Klimov, Paul V. and Quintana, Chris and Acharya, Rajeev and Anderson, Kyle and Ansmann, Markus and Arute, Frank and Arya, Kunal and Asfaw, Abraham and Bardin, Joseph C. and Bourassa, Alexandre and Bovaird, Jenna and Brill, Leon and Buckley, Bob B. and Buell, David A. and Burger, Tim and Burkett, Brian and Bushnell, Nicholas and Campero, Juan and Chiaro, Ben and Collins, Roberto and Conner, Paul and Crook, Alexander L. and Curtin, Ben and Debroy, Dripto M. and Demura, Sean and Dunsworth, Andrew and Erickson, Catherine and Fatemi, Reza and Ferreira, Vinicius S. and Burgos, Leslie Flores and Forati, Ebrahim and Fowler, Austin G. and Foxen, Brooks and Garcia, Gonzalo and Giang, William and Gidney, Craig and Giustina, Marissa and Gosula, Raja and Dau, Alejandro Grajales and Gross, Jonathan
  A. and Hamilton, Michael C. and Harrington, Sean D. and Heu, Paula and Hilton, Jeremy and Hoffmann, Markus R. and Hong, Sabrina and Huang, Trent and Huff, Ashley and Iveland, Justin and Jeffrey, Evan and Jiang, Zhang and Jones, Cody and Kelly, Julian and Kim, Seon and Kostritsa, Fedor and Kreikebaum, John Mark and Landhuis, David and Laptev, Pavel and Laws, Lily and Lee, Kenny and Lester, Brian J. and Lill, Alexander T. and Liu, Wayne and Locharla, Aditya and Lucero, Erik and Martin, Steven and Megrant, Anthony and Mi, Xiao and Montazeri, Shirin and Morvan, Alexis and Naaman, Ofer and Neeley, Matthew and Neill, Charles and Nersisyan, Ani and Newman, Michael and Ng, Jiun How and Nguyen, Anthony and Nguyen, Murray and Potter, Rebecca and Rocque, Charles and Roushan, Pedram and Sankaragomathi, Kannan and Schurkus, Henry F. and Schuster, Christopher and Shearn, Michael J. and Shorter, Aaron and Shutty, Noah and Shvarts, Vladimir and Skruzny, Jindra and Smith, W. Clarke and Sterling, George and Szalay, Marco and
  Thor, Douglas and Torres, Alfredo and White, Theodore and Woo, Bryan W. K. and Yao, Z. Jamie and Yeh, Ping and Yoo, Juhwan and Young, Grayson and Zalcman, Adam and Zhu, Ningfeng and Zobrist, Nicholas and Neven, Hartmut and Smelyanskiy, Vadim and Petukhov, Andre and Korotkov, Alexander N. and Sank, Daniel and Chen, Yu}]{Miao2023a}%
  \BibitemOpen
  \bibfield  {author} {\bibinfo {author} {Miao, K.~C.}, \bibinfo {author} {McEwen, M.}, \bibinfo {author} {Atalaya, J.}, \emph {et~al.},\ }\bibfield  {title} {\bibinfo {title} {Overcoming leakage in quantum error correction},\ }\bibfield  {journal} {\bibinfo  {journal} {Nature Physics}\ }\href {https://doi.org/10.1038/s41567-023-02226-w} {10.1038/s41567-023-02226-w} (\bibinfo {year} {2023})\BibitemShut {NoStop}%
\bibitem [{\citenamefont {Lacroix}\ \emph {et~al.}(2025)Lacroix, Nathan and Hofele, Luca and Remm, Ants and Benhayoune-Khadraoui, Othmane and McDonald, Alexander and Shillito, Ross and Laz\u{a}r, Stefania and Hellings, Christoph and Swiadek, Francois and Colao-Zanuz, Dante and Flasby, Alexander and Panah, Mohsen Bahrami and Kerschbaum, Michael and Norris, Graham J. and Blais, Alexandre and Wallraff, Andreas and Krinner, Sebastian}]{Lacroix2025}%
  \BibitemOpen
  \bibfield  {author} {\bibinfo {author} {Lacroix, N.}, \bibinfo {author} {Hofele, L.}, \bibinfo {author} {Remm, A.}, \emph {et~al.},\ }\bibfield  {title} {\bibinfo {title} {Fast flux-activated leakage reduction for superconducting quantum circuits},\ }\href {https://doi.org/10.1103/PhysRevLett.134.120601} {\bibfield  {journal} {\bibinfo  {journal} {Phys. Rev. Lett.}\ }\textbf {\bibinfo {volume} {134}},\ \bibinfo {pages} {120601} (\bibinfo {year} {2025})}\BibitemShut {NoStop}%
\bibitem [{\citenamefont {Vogel}\ and\ \citenamefont {Risken}(1989)Vogel, K. and Risken, H.}]{Vogel1989}%
  \BibitemOpen
  \bibfield  {author} {\bibinfo {author} {Vogel, K.}\ and\ \bibinfo {author} {Risken, H.},\ }\bibfield  {title} {\bibinfo {title} {Determination of quasiprobability distributions in terms of probability distributions for the rotated quadrature phase},\ }\href {https://doi.org/10.1103/PhysRevA.40.2847} {\bibfield  {journal} {\bibinfo  {journal} {Phys. Rev. A}\ }\textbf {\bibinfo {volume} {40}},\ \bibinfo {pages} {2847} (\bibinfo {year} {1989})}\BibitemShut {NoStop}%
\bibitem [{\citenamefont {Neeley}\ \emph {et~al.}(2010)Neeley, M. and Bialczak, R.~C. and Lenander,~M. and Lucero, E. and Mariantoni, M. and O'Connell, A.~D. and Sank, D. and Wang, H. and Weides, M. and Wenner, J. and Yin, Y. and Yamamoto, T. and Cleland, A.~N. and Martinis, J.~M.}]{Neeley2010a}%
  \BibitemOpen
  \bibfield  {author} {\bibinfo {author} {Neeley, M.}, \bibinfo {author} {Bialczak, R.~C.}, \bibinfo {author} {Lenander, M.}, \emph {et~al.},\ }\bibfield  {title} {\bibinfo {title} {Generation of three-qubit entangled states using superconducting phase qubits},\ }\href {https://doi.org/doi:10.1038/nature09418} {\bibfield  {journal} {\bibinfo  {journal} {Nature}\ }\textbf {\bibinfo {volume} {467}},\ \bibinfo {pages} {570} (\bibinfo {year} {2010})}\BibitemShut {NoStop}%
\bibitem [{\citenamefont {Zhu}(2014)Zhu, Huangjun}]{Zhu2014a}%
  \BibitemOpen
  \bibfield  {author} {\bibinfo {author} {Zhu, H.},\ }\bibfield  {title} {\bibinfo {title} {Quantum state estimation with informationally overcomplete measurements},\ }\href {https://doi.org/10.1103/PhysRevA.90.012115} {\bibfield  {journal} {\bibinfo  {journal} {Phys. Rev. A}\ }\textbf {\bibinfo {volume} {90}},\ \bibinfo {pages} {012115} (\bibinfo {year} {2014})}\BibitemShut {NoStop}%
\bibitem [{\citenamefont {Steffen}\ \emph {et~al.}(2006)Steffen, M. and Ansmann, M. and Bialczak, R. C. and Katz, N. and Lucero, E. and McDermott, R. and Neeley, M. and Weig, E. M. and Cleland, A. N. and Martinis, J. M.}]{Steffen2006a}%
  \BibitemOpen
  \bibfield  {author} {\bibinfo {author} {Steffen, M.}, \bibinfo {author} {Ansmann, M.}, \bibinfo {author} {Bialczak, R.~C.}, \emph {et~al.},\ }\bibfield  {title} {\bibinfo {title} {Measurement of the entanglement of two superconducting qubits via state tomography},\ }\href {http://www.sciencemag.org/content/313/5792/1423.full} {\bibfield  {journal} {\bibinfo  {journal} {Science}\ }\textbf {\bibinfo {volume} {313}},\ \bibinfo {pages} {1423} (\bibinfo {year} {2006})}\BibitemShut {NoStop}%
\bibitem [{\citenamefont {Arunachalam}\ and\ \citenamefont {Schatzki}(2024)Arunachalam, Srinivasan and Schatzki, Louis}]{Arunachalam2024}%
  \BibitemOpen
  \bibfield  {author} {\bibinfo {author} {Arunachalam, S.}\ and\ \bibinfo {author} {Schatzki, L.},\ }\bibfield  {title} {\bibinfo {title} {Distributed inner product estimation with limited quantum communication},\ }\href {https://doi.org/10.48550/ARXIV.2410.12684} {\bibfield  {journal} {\bibinfo  {journal} {arXiv:2410.12684}\ } (\bibinfo {year} {2024})}\BibitemShut {NoStop}%
\bibitem [{\citenamefont {Jozsa}(1994)Richard Jozsa}]{Jozsa1994}%
  \BibitemOpen
  \bibfield  {author} {\bibinfo {author} {Jozsa, R.},\ }\bibfield  {title} {\bibinfo {title} {Fidelity for mixed quantum states},\ }\href {https://doi.org/10.1080/09500349414552171} {\bibfield  {journal} {\bibinfo  {journal} {Journal of Modern Optics}\ }\textbf {\bibinfo {volume} {41}},\ \bibinfo {pages} {2315} (\bibinfo {year} {1994})}\BibitemShut {NoStop}%
\bibitem [{\citenamefont {McKay}\ \emph {et~al.}(2017)McKay, David C. and Wood, Christopher J. and Sheldon, Sarah and Chow, Jerry M. and Gambetta, Jay M.}]{McKay2017}%
  \BibitemOpen
  \bibfield  {author} {\bibinfo {author} {McKay, D.~C.}, \bibinfo {author} {Wood, C.~J.}, \bibinfo {author} {Sheldon, S.}, \emph {et~al.},\ }\bibfield  {title} {\bibinfo {title} {Efficient {$Z$} gates for quantum computing},\ }\href {https://doi.org/10.1103/PhysRevA.96.022330} {\bibfield  {journal} {\bibinfo  {journal} {Phys. Rev. A}\ }\textbf {\bibinfo {volume} {96}},\ \bibinfo {pages} {022330} (\bibinfo {year} {2017})}\BibitemShut {NoStop}%
\bibitem [{\citenamefont {Krinner}\ \emph {et~al.}(2020)Krinner, S. and Laz\u{a}r, S. and Remm, A. and Andersen, C.K. and Lacroix, N. and Norris, G.J. and Hellings, C. and Gabureac, M. and Eichler, C. and Wallraff, A.}]{Krinner2020}%
  \BibitemOpen
  \bibfield  {author} {\bibinfo {author} {Krinner, S.}, \bibinfo {author} {Laz\u{a}r, S.}, \bibinfo {author} {Remm, A.}, \emph {et~al.},\ }\bibfield  {title} {\bibinfo {title} {Benchmarking coherent errors in controlled-phase gates due to spectator qubits},\ }\href {https://doi.org/10.1103/PhysRevApplied.14.024042} {\bibfield  {journal} {\bibinfo  {journal} {Phys. Rev. Appl.}\ }\textbf {\bibinfo {volume} {14}},\ \bibinfo {pages} {024042} (\bibinfo {year} {2020})}\BibitemShut {NoStop}%
\bibitem [{\citenamefont {Elben}\ \emph {et~al.}(2023)Elben, Andreas and Flammia, Steven T. and Huang, Hsin-Yuan and Kueng, Richard and Preskill, John and Vermersch, Beno{\^i}t and Zoller, Peter}]{Elben2022}%
  \BibitemOpen
  \bibfield  {author} {\bibinfo {author} {Elben, A.}, \bibinfo {author} {Flammia, S.~T.}, \bibinfo {author} {Huang, H.-Y.}, \emph {et~al.},\ }\bibfield  {title} {\bibinfo {title} {The randomized measurement toolbox},\ }\href {https://doi.org/10.1038/s42254-022-00535-2} {\bibfield  {journal} {\bibinfo  {journal} {Nature Reviews Physics}\ }\textbf {\bibinfo {volume} {5}},\ \bibinfo {pages} {9} (\bibinfo {year} {2023})}\BibitemShut {NoStop}%
\bibitem [{\citenamefont {H{\"a}ffner}\ \emph {et~al.}(2005)H{\"a}ffner, H. and H{\"a}nsel, W. and Roos, C. F. and Benhelm, J. and Chek-al-kar, D. and Chwalla, M. and Korber, T. and Rapol, U. D. and Riebe, M. and Schmidt, P. O. and Becher, C. and G{\"u}hne, O. and D{\"u}r, W. and Blatt, R.}]{Haeffner2005}%
  \BibitemOpen
  \bibfield  {author} {\bibinfo {author} {H{\"a}ffner, H.}, \bibinfo {author} {H{\"a}nsel, W.}, \bibinfo {author} {Roos, C.~F.}, \emph {et~al.},\ }\bibfield  {title} {\bibinfo {title} {Scalable multiparticle entanglement of trapped ions},\ }\href {http://dx.doi.org/10.1038/nature04279} {\bibfield  {journal} {\bibinfo  {journal} {Nature}\ }\textbf {\bibinfo {volume} {438}},\ \bibinfo {pages} {643} (\bibinfo {year} {2005})}\BibitemShut {NoStop}%
\bibitem [{\citenamefont {Song}\ \emph {et~al.}(2017)Song, Chao and Xu, Kai and Liu, Wuxin and Yang, Chui-ping and Zheng, Shi-Biao and Deng, Hui and Xie, Qiwei and Huang, Keqiang and Guo, Qiujiang and Zhang, Libo and Zhang, Pengfei and Xu, Da and Zheng, Dongning and Zhu, Xiaobo and Wang, H. and Chen, Y.-A. and Lu, C.-Y. and Han, Siyuan and Pan, Jian-Wei}]{Song2017}%
  \BibitemOpen
  \bibfield  {author} {\bibinfo {author} {Song, C.}, \bibinfo {author} {Xu, K.}, \bibinfo {author} {Liu, W.}, \emph {et~al.},\ }\bibfield  {title} {\bibinfo {title} {10-qubit entanglement and parallel logic operations with a superconducting circuit},\ }\href {https://doi.org/10.1103/PhysRevLett.119.180511} {\bibfield  {journal} {\bibinfo  {journal} {Phys. Rev. Lett.}\ }\textbf {\bibinfo {volume} {119}},\ \bibinfo {pages} {180511} (\bibinfo {year} {2017})}\BibitemShut {NoStop}%
\bibitem [{\citenamefont {Gross}\ \emph {et~al.}(2010)Gross, David and Liu, Yi-Kai and Flammia, Steven T. and Becker, Stephen and Eisert, Jens}]{Gross2010a}%
  \BibitemOpen
  \bibfield  {author} {\bibinfo {author} {Gross, D.}, \bibinfo {author} {Liu, Y.-K.}, \bibinfo {author} {Flammia, S.~T.}, \emph {et~al.},\ }\bibfield  {title} {\bibinfo {title} {Quantum state tomography via compressed sensing},\ }\href {https://doi.org/10.1103/physrevlett.105.150401} {\bibfield  {journal} {\bibinfo  {journal} {Physical Review Letters}\ }\textbf {\bibinfo {volume} {105}},\ \bibinfo {pages} {150401} (\bibinfo {year} {2010})}\BibitemShut {NoStop}%
\bibitem [{\citenamefont {Cramer}\ \emph {et~al.}(2010)Cramer, Marcus and Plenio, Martin B. and Flammia, Steven T. and Somma, Rolando and Gross, David and Bartlett, Stephen D. and Landon-Cardinal, Olivier and Poulin, David and Liu, Yi-Kai}]{Cramer2010}%
  \BibitemOpen
  \bibfield  {author} {\bibinfo {author} {Cramer, M.}, \bibinfo {author} {Plenio, M.~B.}, \bibinfo {author} {Flammia, S.~T.}, \emph {et~al.},\ }\bibfield  {title} {\bibinfo {title} {Efficient quantum state tomography},\ }\href {https://doi.org/10.1038/ncomms1147} {\bibfield  {journal} {\bibinfo  {journal} {Nat. Commun.}\ }\textbf {\bibinfo {volume} {1}},\ \bibinfo {pages} {149} (\bibinfo {year} {2010})}\BibitemShut {NoStop}%
\bibitem [{\citenamefont {Acharya}\ \emph {et~al.}(2025{\natexlab{b}})Acharya, J. and Dharmavarapu, A. and Liu, Y. and Yu, N.}]{Acharya2025b}%
  \BibitemOpen
  \bibfield  {author} {\bibinfo {author} {Acharya, J.}, \bibinfo {author} {Dharmavarapu, A.}, \bibinfo {author} {Liu, Y.},\ and\ \bibinfo {author} {Yu, N.},\ }\bibfield  {title} {\bibinfo {title} {Pauli measurements are not optimal for single-copy tomography},\ }\href {https://arxiv.org/abs/2502.18170} {\bibfield  {journal} {\bibinfo  {journal} {arXiv:2502.18170}\ } (\bibinfo {year} {2025}{\natexlab{b}})}\BibitemShut {NoStop}%
\bibitem [{\citenamefont {Binosi}\ \emph {et~al.}(2024)Binosi, D. and Garberoglio, G. and Maragnano, D. and Dapor, M. and Liscidini, M.}]{Binosi2024}%
  \BibitemOpen
  \bibfield  {author} {\bibinfo {author} {Binosi, D.}, \bibinfo {author} {Garberoglio, G.}, \bibinfo {author} {Maragnano, D.}, \emph {et~al.},\ }\bibfield  {title} {\bibinfo {title} {A tailor-made quantum state tomography approach},\ }\href {https://doi.org/10.1063/5.0219143} {\bibfield  {journal} {\bibinfo  {journal} {APL Quantum}\ }\textbf {\bibinfo {volume} {1}},\ \bibinfo {pages} {036112} (\bibinfo {year} {2024})}\BibitemShut {NoStop}%
\bibitem [{\citenamefont {Flammia}\ and\ \citenamefont {Liu}(2011)Flammia, S.~T. and Liu, Y.-K.}]{Flammia2011}%
  \BibitemOpen
  \bibfield  {author} {\bibinfo {author} {Flammia, S.~T.}\ and\ \bibinfo {author} {Liu, Y.-K.},\ }\bibfield  {title} {\bibinfo {title} {Direct fidelity estimation from few pauli measurements},\ }\href {https://doi.org/10.1103/PhysRevLett.106.230501} {\bibfield  {journal} {\bibinfo  {journal} {Phys. Rev. Lett.}\ }\textbf {\bibinfo {volume} {106}},\ \bibinfo {pages} {230501} (\bibinfo {year} {2011})}\BibitemShut {NoStop}%
\bibitem [{\citenamefont {Bartlett}(2014)Bartlett, Stephen D.}]{Bartlett2014}%
  \BibitemOpen
  \bibfield  {author} {\bibinfo {author} {Bartlett, S.~D.},\ }\bibfield  {title} {\bibinfo {title} {Quantum computing: Powered by magic},\ }\href {https://doi.org/doi:10.1038/nature13504} {\bibfield  {journal} {\bibinfo  {journal} {Nature}\ }\textbf {\bibinfo {volume} {510}},\ \bibinfo {pages} {345} (\bibinfo {year} {2014})}\BibitemShut {NoStop}%
\bibitem [{\citenamefont {Huang}\ \emph {et~al.}(2021)Huang, Hsin-Yuan and Kueng, Richard and Preskill, John}]{Huang2021f}%
  \BibitemOpen
  \bibfield  {author} {\bibinfo {author} {Huang, H.-Y.}, \bibinfo {author} {Kueng, R.},\ and\ \bibinfo {author} {Preskill, J.},\ }\bibfield  {title} {\bibinfo {title} {Information-theoretic bounds on quantum advantage in machine learning},\ }\href {https://doi.org/10.1103/PhysRevLett.126.190505} {\bibfield  {journal} {\bibinfo  {journal} {Phys. Rev. Lett.}\ }\textbf {\bibinfo {volume} {126}},\ \bibinfo {pages} {190505} (\bibinfo {year} {2021})}\BibitemShut {NoStop}%
\bibitem [{\citenamefont {Montanaro}(2017)Ashley Montanaro}]{Montanaro2017}%
  \BibitemOpen
  \bibfield  {author} {\bibinfo {author} {Montanaro, A.},\ }\bibfield  {title} {\bibinfo {title} {Learning stabilizer states by bell sampling},\ }\href {https://arxiv.org/abs/1707.04012} {\bibfield  {journal} {\bibinfo  {journal} {arXiv:1707.04012}\ } (\bibinfo {year} {2017})}\BibitemShut {NoStop}%
\bibitem [{\citenamefont {Hangleiter}\ and\ \citenamefont {Gullans}(2024)Hangleiter, Dominik and Gullans, Michael J.}]{Hangleiter2024a}%
  \BibitemOpen
  \bibfield  {author} {\bibinfo {author} {Hangleiter, D.}\ and\ \bibinfo {author} {Gullans, M.~J.},\ }\bibfield  {title} {\bibinfo {title} {Bell sampling from quantum circuits},\ }\bibfield  {journal} {\bibinfo  {journal} {Physical Review Letters}\ }\textbf {\bibinfo {volume} {133}},\ \href {https://doi.org/10.1103/physrevlett.133.020601} {10.1103/physrevlett.133.020601} (\bibinfo {year} {2024})\BibitemShut {NoStop}%
\bibitem [{\citenamefont {Burnett}\ \emph {et~al.}(2019)Burnett, J. and Bengtsson, A. and Scigliuzzo, M. and Niepce, D. and Kudra, M. and Delsing, P. and Bylander, J.}]{Burnett2019}%
  \BibitemOpen
  \bibfield  {author} {\bibinfo {author} {Burnett, J.}, \bibinfo {author} {Bengtsson, A.}, \bibinfo {author} {Scigliuzzo, M.}, \emph {et~al.},\ }\bibfield  {title} {\bibinfo {title} {Decoherence benchmarking of superconducting qubits},\ }\href {https://arxiv.org/abs/1901.04417} {\bibfield  {journal} {\bibinfo  {journal} {npj Quantum Information}\ }\textbf {\bibinfo {volume} {5}},\ \bibinfo {pages} {54} (\bibinfo {year} {2019})}\BibitemShut {NoStop}%
\bibitem [{\citenamefont {Lisenfeld}\ \emph {et~al.}(2023)Lisenfeld, J{\"u}rgen and Bilmes, Alexander and Ustinov, Alexey V.}]{Lisenfeld2023}%
  \BibitemOpen
  \bibfield  {author} {\bibinfo {author} {Lisenfeld, J.}, \bibinfo {author} {Bilmes, A.},\ and\ \bibinfo {author} {Ustinov, A.~V.},\ }\bibfield  {title} {\bibinfo {title} {Enhancing the coherence of superconducting quantum bits with electric fields},\ }\href {https://doi.org/10.1038/s41534-023-00678-9} {\bibfield  {journal} {\bibinfo  {journal} {npj Quantum Information}\ }\textbf {\bibinfo {volume} {9}},\ \bibinfo {pages} {8} (\bibinfo {year} {2023})}\BibitemShut {NoStop}%
\bibitem [{\citenamefont {Krinner}\ \emph {et~al.}(2019)Krinner, S. and Storz, S. and Kurpiers, P. and Magnard, P. and Heinsoo, J. and Keller, R. and L{\"u}tolf, J. and Eichler, C. and Wallraff, A.}]{Krinner2019}%
  \BibitemOpen
  \bibfield  {author} {\bibinfo {author} {Krinner, S.}, \bibinfo {author} {Storz, S.}, \bibinfo {author} {Kurpiers, P.}, \emph {et~al.},\ }\bibfield  {title} {\bibinfo {title} {Engineering cryogenic setups for 100-qubit scale superconducting circuit systems},\ }\href {https://doi.org/10.1140/epjqt/s40507-019-0072-0} {\bibfield  {journal} {\bibinfo  {journal} {EPJ Quantum Technology}\ }\textbf {\bibinfo {volume} {6}},\ \bibinfo {pages} {2} (\bibinfo {year} {2019})}\BibitemShut {NoStop}%
\bibitem [{\citenamefont {Herrmann}\ \emph {et~al.}(2022)Herrmann, J. and Hellings, C. and Laz\u{a}r, S. and Pf\"affli, F. and Haupt, F. and Thiele, T. and Zanuz, D. C. and Norris, G. J. and Heer, F. and Eichler, C. and Wallraff, A.}]{Herrmann2022a}%
  \BibitemOpen
  \bibfield  {author} {\bibinfo {author} {Herrmann, J.}, \bibinfo {author} {Hellings, C.}, \bibinfo {author} {Laz\u{a}r, S.}, \emph {et~al.},\ }\bibfield  {title} {\bibinfo {title} {Frequency up-conversion schemes for controlling superconducting qubits},\ }\href {https://arxiv.org/abs/2210.02513} {\bibfield  {journal} {\bibinfo  {journal} {arXiv:2210.02513}\ } (\bibinfo {year} {2022})}\BibitemShut {NoStop}%
\bibitem [{\citenamefont {Macklin}\ \emph {et~al.}(2015)Macklin, C. and O'Brien, K. and Hover, D. and Schwartz, M. E. and Bolkhovsky, V. and Zhang, X. and Oliver, W. D. and Siddiqi, I.}]{macklin2015}%
  \BibitemOpen
  \bibfield  {author} {\bibinfo {author} {Macklin, C.}, \bibinfo {author} {O'Brien, K.}, \bibinfo {author} {Hover, D.}, \emph {et~al.},\ }\bibfield  {title} {\bibinfo {title} {A near-quantum-limited {Josephson} traveling-wave parametric amplifier},\ }\href {https://doi.org/10.1126/science.aaa8525} {\bibfield  {journal} {\bibinfo  {journal} {Science}\ }\textbf {\bibinfo {volume} {350}},\ \bibinfo {pages} {307} (\bibinfo {year} {2015})}\BibitemShut {NoStop}%
\bibitem [{\citenamefont {Bultink}\ \emph {et~al.}(2018)C. C. Bultink and B. Tarasinski and N. Haandb{\ae}k and S. Poletto and N. Haider and D. J. Michalak and A. Bruno and L. DiCarlo}]{Bultink2018}%
  \BibitemOpen
  \bibfield  {author} {\bibinfo {author} {Bultink, C.~C.}, \bibinfo {author} {Tarasinski, B.}, \bibinfo {author} {Haandb{\ae}k, N.}, \emph {et~al.},\ }\bibfield  {title} {\bibinfo {title} {General method for extracting the quantum efficiency of dispersive qubit readout in circuit {QED}},\ }\href {https://doi.org/10.1063/1.5015954} {\bibfield  {journal} {\bibinfo  {journal} {Appl. Phys. Lett.}\ }\textbf {\bibinfo {volume} {112}},\ \bibinfo {pages} {092601} (\bibinfo {year} {2018})}\BibitemShut {NoStop}%
\bibitem [{\citenamefont {Javadi-Abhari}\ \emph {et~al.}(2024)Javadi-Abhari, Ali and Treinish, Matthew and Krsulich, Kevin and Wood, Christopher J. and Lishman, Jake and Gacon, Julien and Martiel, Simon and Nation, Paul D. and Bishop, Lev S. and Cross, Andrew W. and Johnson, Blake R. and Gambetta, Jay M.}]{Javadi-Abhari2024}%
  \BibitemOpen
  \bibfield  {author} {\bibinfo {author} {Javadi-Abhari, A.}, \bibinfo {author} {Treinish, M.}, \bibinfo {author} {Krsulich, K.}, \emph {et~al.},\ }\bibfield  {title} {\bibinfo {title} {Quantum computing with qiskit},\ }\href {http://arxiv.org/abs/2405.08810} {\bibfield  {journal} {\bibinfo  {journal} {arXiv:2405.08810}\ } (\bibinfo {year} {2024})}\BibitemShut {NoStop}%
\bibitem [{\citenamefont {Chen}\ \emph {et~al.}(2021)Chen, Zijun and Satzinger, Kevin J. and Atalaya, Juan and Korotkov, Alexander N. and Dunsworth, Andrew and Sank, Daniel and Quintana, Chris and McEwen, Matt and Barends, Rami and Klimov, Paul V. and Hong, Sabrina and Jones, Cody and Petukhov, Andre and Kafri, Dvir and Demura, Sean and Burkett, Brian and Gidney, Craig and Fowler, Austin G. and Paler, Alexandru and Putterman, Harald and Aleiner, Igor and Arute, Frank and Arya, Kunal and Babbush, Ryan and Bardin, Joseph C. and Bengtsson, Andreas and Bourassa, Alexandre and Broughton, Michael and Buckley, Bob B. and Buell, David A. and Bushnell, Nicholas and Chiaro, Benjamin and Collins, Roberto and Courtney, William and Derk, Alan R. and Eppens, Daniel and Erickson, Catherine and Farhi, Edward and Foxen, Brooks and Giustina, Marissa and Greene, Ami and Gross, Jonathan A. and Harrigan, Matthew P. and Harrington, Sean D. and Hilton, Jeremy and Ho, Alan and Huang, Trent and Huggins, William J. and Ioffe, L. B.
  and Isakov, Sergei V. and Jeffrey, Evan and Jiang, Zhang and Kechedzhi, Kostyantyn and Kim, Seon and Kitaev, Alexei and Kostritsa, Fedor and Landhuis, David and Laptev, Pavel and Lucero, Erik and Martin, Orion and McClean, Jarrod R. and McCourt, Trevor and Mi, Xiao and Miao, Kevin C. and Mohseni, Masoud and Montazeri, Shirin and Mruczkiewicz, Wojciech and Mutus, Josh and Naaman, Ofer and Neeley, Matthew and Neill, Charles and Newman, Michael and Niu, Murphy Yuezhen and O’Brien, Thomas E. and Opremcak, Alex and Ostby, Eric and Pató, Bálint and Redd, Nicholas and Roushan, Pedram and Rubin, Nicholas C. and Shvarts, Vladimir and Strain, Doug and Szalay, Marco and Trevithick, Matthew D. and Villalonga, Benjamin and White, Theodore and Yao, Z. Jamie and Yeh, Ping and Yoo, Juhwan and Zalcman, Adam and Neven, Hartmut and Boixo, Sergio and Smelyanskiy, Vadim and Chen, Yu and Megrant, Anthony and Kelly, Julian}]{Chen2021p}%
  \BibitemOpen
  \bibfield  {author} {\bibinfo {author} {Chen, Z.}, \bibinfo {author} {Satzinger, K.~J.}, \bibinfo {author} {Atalaya, J.}, \emph {et~al.},\ }\bibfield  {title} {\bibinfo {title} {Exponential suppression of bit or phase errors with cyclic error correction},\ }\href {https://doi.org/10.1038/s41586-021-03588-y} {\bibfield  {journal} {\bibinfo  {journal} {Nature}\ }\textbf {\bibinfo {volume} {595}},\ \bibinfo {pages} {383} (\bibinfo {year} {2021})}\BibitemShut {NoStop}%
\bibitem [{\citenamefont {Aaronson}(2018)Aaronson, Scott}]{Aaronson2018}%
  \BibitemOpen
  \bibfield  {author} {\bibinfo {author} {Aaronson, S.},\ }\bibfield  {title} {\bibinfo {title} {Shadow tomography of quantum states},\ }in\ \href {https://doi.org/10.1145/3188745.3188802} {\emph {\bibinfo {booktitle} {Proceedings of the 50th Annual ACM SIGACT Symposium on Theory of Computing}}},\ \bibinfo {series and number} {STOC ’18}\ (\bibinfo  {publisher} {ACM},\ \bibinfo {year} {2018})\ pp.\ \bibinfo {pages} {325--338}\BibitemShut {NoStop}%
\bibitem [{\citenamefont {Raginsky}(2001)Maxim Raginsky}]{Raginsky2001}%
  \BibitemOpen
  \bibfield  {author} {\bibinfo {author} {Raginsky, M.},\ }\bibfield  {title} {\bibinfo {title} {A fidelity measure for quantum channels},\ }\href {https://doi.org/https://doi.org/10.1016/S0375-9601(01)00640-5} {\bibfield  {journal} {\bibinfo  {journal} {Physics Letters A}\ }\textbf {\bibinfo {volume} {290}},\ \bibinfo {pages} {11} (\bibinfo {year} {2001})}\BibitemShut {NoStop}%
\bibitem [{\citenamefont {Dankert}\ \emph {et~al.}(2009)Dankert, Christoph and Cleve, Richard and Emerson, Joseph and Livine, Etera}]{Dankert2009}%
  \BibitemOpen
  \bibfield  {author} {\bibinfo {author} {Dankert, C.}, \bibinfo {author} {Cleve, R.}, \bibinfo {author} {Emerson, J.},\ and\ \bibinfo {author} {Livine, E.},\ }\bibfield  {title} {\bibinfo {title} {Exact and approximate unitary 2-designs and their application to fidelity estimation},\ }\href {https://doi.org/10.1103/PhysRevA.80.012304} {\bibfield  {journal} {\bibinfo  {journal} {Phys. Rev. A}\ }\textbf {\bibinfo {volume} {80}},\ \bibinfo {pages} {012304} (\bibinfo {year} {2009})}\BibitemShut {NoStop}%
\bibitem [{\citenamefont {Gross}\ \emph {et~al.}(2007)Gross, D. and Audenaert, K. and Eisert, J.}]{Gross2007}%
  \BibitemOpen
  \bibfield  {author} {\bibinfo {author} {Gross, D.}, \bibinfo {author} {Audenaert, K.},\ and\ \bibinfo {author} {Eisert, J.},\ }\bibfield  {title} {\bibinfo {title} {Evenly distributed unitaries: On the structure of unitary designs},\ }\bibfield  {journal} {\bibinfo  {journal} {Journal of Mathematical Physics}\ }\textbf {\bibinfo {volume} {48}},\ \href {https://doi.org/10.1063/1.2716992} {10.1063/1.2716992} (\bibinfo {year} {2007})\BibitemShut {NoStop}%
\bibitem [{\citenamefont {Brand{\~a}o}\ \emph {et~al.}(2016)Brand{\~a}o, Fernando G. S. L. and Harrow, Aram W. and Horodecki, Micha{\l}}]{Brandao2016c}%
  \BibitemOpen
  \bibfield  {author} {\bibinfo {author} {Brand{\~a}o, F. G. S.~L.}, \bibinfo {author} {Harrow, A.~W.},\ and\ \bibinfo {author} {Horodecki, M.},\ }\bibfield  {title} {\bibinfo {title} {Local random quantum circuits are approximate polynomial-designs},\ }\href {https://doi.org/10.1007/s00220-016-2706-8} {\bibfield  {journal} {\bibinfo  {journal} {Communications in Mathematical Physics}\ }\textbf {\bibinfo {volume} {346}},\ \bibinfo {pages} {397} (\bibinfo {year} {2016})}\BibitemShut {NoStop}%
\bibitem [{\citenamefont {Horodecki}\ \emph {et~al.}(1999)Horodecki, M. and Horodecki, P. and Horodecki, R.}]{Horodecki1999}%
  \BibitemOpen
  \bibfield  {author} {\bibinfo {author} {Horodecki, M.}, \bibinfo {author} {Horodecki, P.},\ and\ \bibinfo {author} {Horodecki, R.},\ }\bibfield  {title} {\bibinfo {title} {General teleportation channel, singlet fraction, and quasidistillation},\ }\href {http://pra.aps.org/abstract/PRA/v60/i3/p1888_1} {\bibfield  {journal} {\bibinfo  {journal} {Phys. Rev. A}\ }\textbf {\bibinfo {volume} {60}},\ \bibinfo {pages} {1888} (\bibinfo {year} {1999})}\BibitemShut {NoStop}%
\bibitem [{\citenamefont {Nielsen}(2002)Nielsen, Michael A.}]{Nielsen2002}%
  \BibitemOpen
  \bibfield  {author} {\bibinfo {author} {Nielsen, M.~A.},\ }\bibfield  {title} {\bibinfo {title} {A simple formula for the average gate fidelity of a quantum dynamical operation},\ }\href {https://www.sciencedirect.com/science/article/pii/S0375960102012720} {\bibfield  {journal} {\bibinfo  {journal} {Phys. Lett. A}\ }\textbf {\bibinfo {volume} {303}},\ \bibinfo {pages} {249} (\bibinfo {year} {2002})}\BibitemShut {NoStop}%
\end{thebibliography}%

\end{document}